\documentclass[twocolumn]{aastex631}

\usepackage{ulem}
\usepackage{xcolor}
\usepackage{graphicx}   % Including figure files
\usepackage{amsmath}    % Advanced maths commands
\usepackage{amssymb}    % Extra maths symbols
\usepackage{diagbox}    % Extra maths symbols
\usepackage{subfigure}
\usepackage{float}
\usepackage{booktabs}
\usepackage{threeparttable}
\usepackage{color}
\usepackage{subfigure}
\usepackage{cancel}
\usepackage{soul}
\usepackage{graphicx}
\usepackage{float}
\usepackage{natbib}
\usepackage{dcolumn}% Align table columns on decimal point
\usepackage{bm}% bold math
\usepackage{ulem}
\usepackage{color}
\usepackage{txfonts}
\usepackage{rotating}
\usepackage{booktabs}
\usepackage{array}
\usepackage{tabularx}
\usepackage{threeparttable}
\usepackage{diagbox}
\usepackage{multirow}
\usepackage{makecell}
\usepackage{longtable}

\shorttitle{The population synthesis of Wolf-Rayet stars involving binary merger channels}
\shortauthors{Zhuowen Li et al.}

\begin{document}

\title{The population synthesis of Wolf-Rayet stars involving binary merger channels}

\correspondingauthor{Chunhua Zhu, Guoliang L\"{u}}
\email{chunhuazhu@sina.cn, guolianglv@xao.ac.cn}

\author[0009-0006-1716-357X]{Zhuowen Li}
\email{ZhuoWenli2024@163.com}
\affiliation{School of Physical Science and Technology, Xinjiang University, Urumqi, 830046, China}

\author{Chunhua Zhu}
\affiliation{School of Physical Science and Technology, Xinjiang University, Urumqi, 830046, China}

\author{Guoliang L\"{u}}
\affil{Xinjiang Observatory,
the Chinese Academy of Sciences, Urumqi, 830011, China}
\affil{School of Physical Science and Technology,
Xinjiang University, Urumqi, 830046, China}

\author{Lin Li}
\affiliation{School of Physical Science and Technology, Xinjiang University, Urumqi, 830046, China}

\author{Helei Liu}
\affiliation{School of Physical Science and Technology, Xinjiang University, Urumqi, 830046, China}

\author{Sufen Guo}
\affiliation{School of Physical Science and Technology, Xinjiang University, Urumqi, 830046, China}

\author[0000-0001-8493-5206]{Jinlong Yu}
\affiliation{College of Mechanical and Electronic Engineering, Tarim University, Alar, 843300 China}

\author[0000-0002-3849-8962]{Xizhen Lu}
\affiliation{School of Physical Science and Technology, Xinjiang University, Urumqi, 830046, China}

\begin{abstract}
Wolf-Rayet stars (WRs) are very important massive stars. However, their origin and the observed binary fraction within the entire WR population are still debated. We investigate some possible merger channels for the formation of WRs, including main sequence (MS)/ Hertzsprung Gap (HG) + MS, He + HG/ Giant Branch (GB). We find that many products produced via binary merger can evolve into WRs,  the MS/ HG + MS merger channel can explain WRs with luminosities higher than $\sim 10^{5.4}$\,L$_{\odot}$, while the He + HG/ GB merger channel can explain low-luminosity WRs in the range of $10^{4.7}$\,L$_{\odot}$\,$\sim$\,$10^{5.5}$\,L$_{\odot}$. In the population synthesis analysis of WRs, we assume an initial binary fraction ($f_{\rm ini,bin}$) of 50\% and 100\% for massive stars. We also assume that MS/ HG + MS merger products are non-rotating or rapidly rotating ($\omega/\omega_{\rm crit}=0.8$). In different cases, the calculated single fractions of WRs range from $22.2\%$ to $60.6\%$ in the Milky Way (MW) and from $8.3\%$ to $70.9\%$ in the Large Magellanic Cloud (LMC). The current observations fall within the range of our calculations. When the merger product of MS/HG+MS rotates rapidly, we estimate that there are approximately 1015 to 1396 WRs in the MW and 128 to 204 WRs in the LMC. Our model also roughly reproduces the observed single-peak luminosity distribution of WRs in the MW. However, the weak bimodal luminosity distribution observed in the LMC is not reproduced in our model. We assess that this may be due to the model underestimating the mass-loss rate in the LMC. In conclusion, we consider that the binary merger is significant formation channel for WR formation, and can explain the observed high fraction of the single WRs in the total population.
\end{abstract}


\section{Introduction}

Wolf-Rayet stars (WRs) are typically massive helium-burning stars that have lost their hydrogen envelopes. It is characterized by broad and intense emission lines \citep{2007ARA&A..45..177C}. The strong stellar winds of WRs contribute to the nucleosynthesis of elements and drive the chemical evolution of galaxies \citep{2003ApJ...591..288H,2012ARA&A..50..107L}. Moreover, the study of WRs is crucial for gaining insights into various important objects, such as X-ray binaries and black holes \citep{2017MNRAS.471.4256V}.

The origin of WRs is still poorly understood, despite their significance and the progress made in studying them. Some calculations using single-star evolutionary models can explain a significant portion of the properties of highly luminous WRs through high mass-loss rate caused by strong stellar winds and/ or rotation induced mixing \citep{1994A&A...290..819L,2005A&A...429..581M,2018PASP..130h4202C}. However, the mass-loss rate ($\dot{M}_{\rm loss}$) of massive stars remains highly uncertain \citep{2011A&A...526A.156M}. Additionally, observations indicate that most massive early O/B-type stars do not exhibit very high rotation \citep{2013A&A...560A..29R,2015A&A...580A..92R,2017A&A...600A..81R}.

Theoretically, binary evolutionary models provide a more robust explanation for the formation of WRs. In these models, the donor evolves into WRs by losing most of its H-rich envelope during the mass transfer (MT) process \citep{2019A&A...627A.151S}. \cite{2022A&A...667A..58P} also speculate that 80\% of the WR population in the Large Magellanic Cloud (LMC) consists of binary systems. However, observationally, many WRs do not have identified companions, and the fraction of single WRs to the entire WR population is estimated to be over 60\% in the Milky Way (MW) \citep{1980A&A....88..230V,1998A&ARv...9...63V,2001NewAR..45..135V,2020A&A...641A..26D,2022A&A...664A..93D}, and around 70\% for the LMC \citep{2001MNRAS.324...18B,2003MNRAS.338..360F,2003MNRAS.338.1025F,2008MNRAS.389..806S,2014ApJ...789...10N,2014A&A...565A..27H,2019A&A...627A.151S}. Additionally, the optically thick wind effect in H-free WRs leads to observed effective temperatures that are much lower than model calculations, posing challenges for theoretical models in explaining H-free WRs \citep{2012A&A...538A..40G,2023A&A...674A.216L}.

It is well known that more than 50\% of massive stars are born in binary systems \citep{2013A&A...550A.107S}. Additionally, over 30\% of the massive stars that exchange mass with a companion will merge \citep{2012Sci...337..444S}. Therefore, it is interesting to investigate whether massive binary mergers are the main formation scenario for WRs and what properties these WRs possess. In this paper, we study the WR population resulting from binary merger channels. Our paper is structured as follows: in Section 2, we describe the modeling of different merging channels, the evolutionary trajectories, and the properties of the resulting WRs. In Section 3, we simulate the WR population using a synthetic population method. Finally, in Section 4, we present our conclusions.

\section{WRs produced via binary merger}

In this section, we provide details of the modelling of the different merging channels and compare the evolutionary trajectories of the merger products from these channels with the observed positions of WRs in the Hertzsprung-Russell diagram (HRD).

\subsection{Model}

Binary mergers are complex and challenging to simulate accurately. To simplify the process, we utilize the rapid binary code of Massive Objects in Binary Stellar Evolution (MOBSE) \citep{2000MNRAS.315..543H,2002MNRAS.329..897H,2018MNRAS.474.2959G} to calculate the binary evolution leading up to the merger event. During this evolution, the dynamical MT between the binary components can result in a common envelope evolution (CEE), ultimately leading to the formation of a close binary or direct merger into a new star \cite[eg., ][]{2020RAA....20..161H}.

In the MOBSE code, the evolution and structure of stars are described by a series of fitting formulae, which are based on calculations using stellar models from \cite{1998MNRAS.298..525P} \citep{2000MNRAS.315..543H,2002MNRAS.329..897H,2006MNRAS.369.1152K,2018MNRAS.474.2959G}. The outcomes of binary system interactions are mainly determined by parameters for common envelope evolution (CEE) and the mass transfer (MT) model. During the CE phase, MOBSE uses a standard energy prescription for treatment \citep{2002MNRAS.329..897H,2013A&ARv..21...59I,2018MNRAS.474.2959G}. The efficiency for the orbital energy used to eject the envelope is set to $\alpha_{\rm CE}=1$, while the envelope structure parameter is $\lambda=0.1$ \citep{2010ApJ...716..114X}. Additionally, the MT model and any unmentioned parameter settings align with the default model (values) outlined in \cite{2018MNRAS.474.2959G}.

Based on the studies by \cite{2002MNRAS.329..897H} and \cite{2018MNRAS.474.2959G}, we consider the following merger scenarios for the formation and evolution of WRs: merger between a main sequence (MS)/Hertzsprung Gap (HG) star and a MS star, and merger between a helium star (He) and a HG/ Giant Branch (GB). The structure and evolution of the newly formed star resulting from the binary merger are simulated using the Modules for Experiments in Stellar Astrophysics (MESA) code (version 10398) \citep{2011ApJS..192....3P,2013ApJS..208....4P,2015ApJS..220...15P,2018ApJS..234...34P}. The subsequent subsections provide detailed explanations of these merger channels. In addition, we use MESA to cross-check the MOBSE calculations. We try to keep the two codes as consistent as possible. If there is an inconsistency between the results of MOBSE and MESA calculations, we use the results calculated by MESA. In MESA, based on \cite{2009MNRAS.396.1699S}, \cite{2021ApJ...920...81S}, \cite{2022A&A...667A..58P}, and \cite{2023arXiv231212658D}, for binary systems with very short initial orbital periods, the donor undergoes very efficient mass stripping while the gainer rapidly expands and reaches critical rotation because it accretes too much material within a short timescale. The gainer that reaches critical rotation is usually considered not to continue accreting material, and excess material is lost from the vicinity of the gainer \citep{2009MNRAS.396.1699S,2021ApJ...920...81S}. In our simulations, we find that most binary systems with initial orbit periods of about $\le4$\,days, the gainer expands significantly and reaches critical rotation during MT. Only a few binary systems have the gainer rotation speeds that may stay below the critical value due to tidal synchronization \citep{1988A&A...202...93R}. Following \cite{1994inbi.conf..263V} and \cite{2024MNRAS.529.1886H}, we use a parameter $\beta=0.75$ ($\beta$ is the fraction of mass lost from the vicinity of the gainer) for this case. For binary systems with initial orbital periods of about $>4$ days, we find that the mass stripping efficiency of the donor is relatively weak, and conservative MT can achieve MT occurring in case A/B.

\subsubsection{MS/HG + MS merger model}

In the case of massive binaries with small orbital periods and/ or large eccentricity, the merger occurs when the two parent stars are brought into deeper contact. This is caused by the continued expansion of the primary as it evolves into the MS/ HG phase, as well as the sharp orbital contraction resulting from the loss of angular momentum \citep{1992ApJ...391..246P,2001A&A...369..939W,2014ApJ...782....7D}. The merger product has a higher mass and luminosity compared to the two parent stars (the primary and secondary before the merger) \citep{2013MNRAS.434.3497G}, which in turn drives stronger stellar winds \citep{2001A&A...369..574V}. This implies the possibility of the formation of WRs in the subsequent evolution of these merger products.

In the MOBSE code, the two stars in the binary system will merge if the condition $(R_1 + R_2) > a(1-e)$ \citep{2002MNRAS.329..897H,2017PASA...34...58E,2018MNRAS.474.2959G,2022ApJS..258...34R} is met during the expansion of the MS/ HG phase. Here, $a$ represents the binary separation, $e$ is the eccentricity, and $R_1$ and $R_2$ are the radii of the two components. Following the approach of \cite{1997MNRAS.291..732T} and \cite{2002MNRAS.329..897H}, the material of the two merging components is assumed to be completely mixed. The hydrogen in the core of the merger product will be replenished, resulting in the rejuvenation of the merged star to approximately a zero-age main sequence (ZAMS) star \citep{1983Ap&SS..96...37H,1998A&ARv...9...63V,2021MNRAS.507.5013M}. Due to the relatively large binding energy of the MS/ HG stars \citep{2000ARA&A..38..113T,2004ApJ...601.1058I,2012ApJ...759...52D}, no mass loss is assumed during the collision. In other words, the mass of the merger product is equal to the sum of the masses of the two parent stars, i.e., $M_{\text{merger}} = M_1 + M_2$.

The post-merge evolution of the merger product is performed using the MESA code. For the simulations, we adopt typical metallicities of 0.014 for the MW \citep{2012A&A...537A.146E} and 0.006 for the LMC \citep{2021A&A...652A.137E}, as WRs are mainly found in the MW and the LMC. In the MESA simulations, convection is treated using the Ledoux criterion and a mixing length parameter of 1.5 \citep{1991A&A...252..669L}. A semi-convective parameter of 1 is also used \citep{1991A&A...252..669L}. The overshoot area is calculated using a step-function approach, with an overshooting parameter set to 0.335 \citep{2011A&A...530A.115B}. Additionally, a thermohaline mixing parameter of 1 is employed \citep{1980A&A....91..175K}. To avoid computational convergence failure of the model near the Eddington limit, we enable the option MLT++ \citep{2013ApJS..208....4P}.

Mass loss is calculated with reference to \cite{2011A&A...530A.115B} and \cite{2022A&A...667A..58P}. When the temperature is below the bi-stability jump temperature, the maximum value of the mass-loss rate from \cite{2001A&A...369..574V} or \cite{1990A&A...231..134N} is used. When the temperature is higher than the bi-stability jump temperature, the mass-loss rate for MS stars is calculated using the prescription of \cite{2001A&A...369..574V}. For the surface H abundance ($X_{\rm H}$) is between $10^{-5}$ and 0.4, the mass-loss rate from \cite{2000A&A...360..227N} is used. For the H-free WR stage ($X_{\rm H} < 10^{-5}$), the mass-loss rate from \cite{2017MNRAS.470.3970Y} is employed. For $X_{\rm H}$ is between 0.4 and 0.7, the wind mass-loss rate is linearly interpolated between the prescriptions of \cite{2001A&A...369..574V} and \cite{2000A&A...360..227N}. Since the work of \cite{2006ApJ...637.1025F}, \cite{2006A&A...454..625P}, \cite{2017A&A...606A..31K}, \cite{2020MNRAS.492.5994B}, and \cite{2021A&A...648A..36B} suggests that the mass-loss rate of massive stars at different stages is not as high as expected, we adopt a similar approach to \cite{2006A&A...460..199Y} and \cite{2011A&A...530A.115B}, reducing the mass-loss rate from \cite{2000A&A...360..227N} by a factor of 10. It is worth noting that we use the same stellar wind scheme described above for both the MW and LMC. Considering that the MW has a higher mass-loss rate, while the LMC has a lower one, our stellar wind model may be reasonable for the LMC, while it may be underestimated for the MW.

In the post-merge evolution, the effects of angular momentum conservation, off-center collisions \citep{2022MNRAS.516.1072C}, and the formation of an excretion disk around the merged product \citep{2005ApJ...623..302F} are taken into consideration. These complexities can result in the merged product either rotating or not rotating. To quantify the effect of rotation on the merger product, a ratio of the rotational angular velocity to the critical rotational angular velocity, $\omega / \omega_{\rm crit}$, is used. In the simulations, values of $\omega / \omega_{\rm crit} = 0$ and $0.8$ are adopted. The efficiency parameter for rotational mixing is set to $1/30$ \citep{2011A&A...530A.115B}, and the inhibition parameter for the chemical gradient is set to 0.1 \citep{2011A&A...530A.115B}. We also considered various instabilities caused by rotation, such as dynamical shear instability, secular shear instability, Eddington-Sweet circulation, and the Goldreich-Schubert-Fricke instability \citep{2000ApJ...528..368H}. Furthermore, the mass-loss rate of a rotating massive star is increased according to the equation $\dot{M} = \left(\frac{1}{1-\omega / \omega_{\rm crit}}\right)^{\beta} \dot{M}_{v_{\rm rot}=0}$, where $\beta=0.43$ \citep{1998A&A...329..551L}.

\subsubsection{He + HG/ GB merger model}

The binary systems including a He and a HG or GB star generally have undergone a stable MT or a CEE. Its companion can fill its Roche lobe at a subsequent evolutionary stage (HG/GB). In MOBSE, the He star and the HG or GB star experience a CEE if $q=\frac{M_{\rm donor}}{M_{\rm gainer}} > q_{\rm crit}$. For a HG star as the donor: $q_{\rm crit,HG}=4$ \citep{1997MNRAS.291..732T}; for a GB star as the donor: $q_{\rm crit,GB}=0.362+[3(1-M_{2,c}/M_{2})]^{-1}$ \citep{1987ApJ...318..794H},
where $M_{2,c}$ is the core mass of donor. Based on a standard CEE model, He star and its companion can merge into a new star if $E_{bind}>\alpha_{\rm CE}\Delta E_{\rm orb}$, where $\Delta E_{\rm orb}$ is the change of orbital energy of the binary from CEE begin to end, $E_{\rm bind}$ is the binding energy of the donor envelope, and $\alpha_{\rm CE}$ is the efficiency for the orbital energy  used to expel the envelope \citep{2002MNRAS.329..897H,2013A&ARv..21...59I}. Based on \cite{1983ApJ...268..368E}, \cite{1995MNRAS.272..800H}, \cite{1997MNRAS.291..732T}, \cite{2010ApJ...716..114X} and \cite{2020MNRAS.495.4659F}, the mass lost ($\Delta m$) of the donor due to the orbital energy changes ($\Delta E_{\rm orb}$) during CEE can be approximately calculated by $\int_{M_{2,c}}^{M_{2,env}-\Delta m}\left(-\frac{G m}{r}+\alpha_{\rm th} u\right) d m=\alpha_{\rm CE} \Delta E_{\rm orb}$, where $M_{\rm 2,env}$ is the envelope mass of the donor, $u$ is the specific internal energy of the gas, the thermal ejection efficiency parameter $\alpha_{\rm th}$ is 1, $\alpha_{\rm CE}$ is 1, the core boundary is defined as the central H-poor region of $X_{\rm H}<10^{-4}$.

For the structure of the merged product, following \cite{2002ApJ...568..939L} and \cite{2008MNRAS.383L...5G}, the He star with higher density (lower entropy) sinks into the core of the donor as the core of the merged product. According to \cite{2013MNRAS.430.2113Z} and \cite{2020ApJ...889...33Z}, the merging process is approximated by He stars accreting the average mass fraction of the donor core initially, followed by accreting the average mass fraction of the remaining envelope ($M_{\rm env}-\Delta m$), with an accretion rate ($\dot M_{\rm acc}$) of 1\,M$_{\odot}$/yr. That is, the new born star has a core mass combining He star and the donor core, and a H-rich envelope mass of $M_{\rm env}-\Delta m$.

\subsection{Evolutionary Tracks of WRs produced via binary merger}
Our study focuses on the evolution of the merger products and explores whether these products from different merger channels can evolve into WRs. We aim to determine which subtypes of WRs can be explained by these different merger channels.

Based on previous studies by \cite{1991A&A...241...77S}, \cite{2003A&A...404..975M} and \cite{2012A&A...542A..29G}, the different stages or subtypes of stars in our evolutionary model are defined as follows: He stars ($\log(T_{\rm eff}/{\rm K})>4.0$ and $X_{\rm H}<0.3$), WRs ($\log(T_{\rm eff}/{\rm K})>4.0$, $X_{\rm H}<0.3$, and the mass-loss rate $>10^{-6}$ M$_{\odot}/{\rm yr}$) \citep{2007ARA&A..45..177C}. H-rich WNs ($10^{-5}<X_{\rm H}<0.3$), H-free WNs ($X_{\rm H}<10^{-5}$ and $X_{\rm N}>X_{\rm C}$), and WCs ($X_{\rm H}<10^{-5}$ and $X_{\rm N}<X_{\rm C}$). \cite{2022A&A...661A..60A} point out that WOs may be special WCs, so possible WOs phase is not considered in the model. When considering the observed sample of WNs, we follow the distinction made by \cite{2014A&A...565A..27H} and \cite{2022A&A...667A..58P}, where WNs with $X_{\rm H}>0$ are classified as H-rich WNs and those with $X_{\rm H}=0$ are classified as H-free WNs. In addition, there are some H-rich WNs with $X_{\rm H}>0.3$ in the observed WRs of \cite{2014A&A...565A..27H} and \cite{2019A&A...625A..57H}. For example, $X_{\rm H}$ of BAT99-49 and BAT99-111, their $X_{\rm H}$ may be higher than 0.5. However, \cite{2014A&A...565A..27H} and \cite{2022A&A...667A..58P} considered that these stars are unlikely to be WRs but are instead more likely to be central hydrogen-burning stars. Considering that the $X_{\rm H}$ values of these samples exceed the range designated for WRs in the studies by \cite{1991A&A...241...77S}, \cite{2003A&A...404..975M}, \cite{2012A&A...542A..29G}, and \cite{2013MNRAS.433.1114Y}, we exclude these observed samples from our analysis. Furthermore, some WCs observed in the LMC lack information about their effective temperature. Consequently, this subset of the sample is excluded from our HRD analysis.

\subsubsection{Evolution of MS/ HG + MS mergers}
For the MS/MS+HG merger product, based on \cite{2016MNRAS.457.2355S}, hydrogen can mix into the center of the merger product during the merger process, serving as fuel. The core mass and luminosity of the merger product are larger and higher than those of its two parent stars (the primary and secondary before the merger). In our model, we assume that the merger product rejuvenates to become a ZAMS star. In our model, we assume that the merger product is rejuvenated to a ZAMS star.

\begin{figure*}[htb]
\centering
\includegraphics[width=\textwidth]{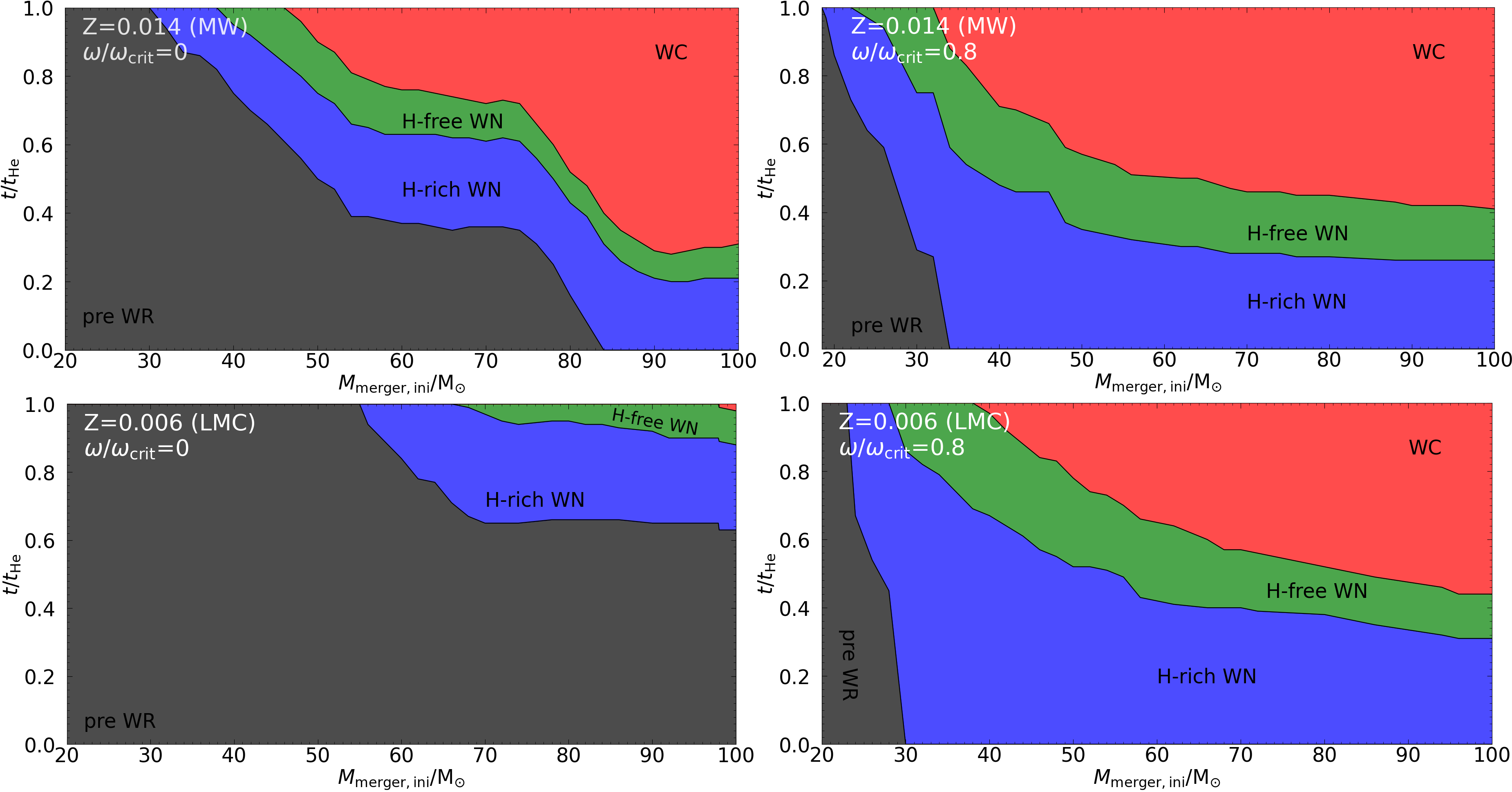}
\caption{MS/HG + MS merger products exhibit different WRs phases during post-MS evolution. These phases, including the pre WRs phase (black), H-rich WNs phase (blue), H-free WNs phase (green), and WCs phase (red), can be observed as a function of the initial mass of the merger products, the normalized time to He burning, and variations in metallicities and rotations.}
\label{fig:1}
\end{figure*}

Fig. \ref{fig:1} presents the evolutionary stage of the merger product as a function of its initial mass and the normalized time during the post-MS evolution. Rotation lowers the lower limit of the initial mass of the merger product for the formation of H-rich WNs, H-free WNs, and WCs, attributed to the increased mass-loss rate of merger products due to rotation \citep{1998A&A...329..551L}. In the MW, without rotation, a merger product with mass higher than approximately 30\,M$_{\odot}$ can evolve into the H-rich WNs phase. However, the lower mass limits in models with a rapid rotation of 0.8\,$\omega_{\rm crit}$ are around 19\,M$_{\odot}$. Without rotation, a merger product with mass of about 84\,M$_{\odot}$ can enter the H-rich WNs phase at the terminal-age main sequence (TAMS). However, in models with a rapid rotation of 0.8\,$\omega_{\rm crit}$, this mass requirement drops to around 34\,M$_{\odot}$. In the LMC, without rotation, the merger product needs approximately 55\,M$_{\odot}$ to evolve into the H-rich WNs phase. With a rapid rotational velocity of 0.8\,$\omega_{\rm crit}$, this lower limit decreases to about 24\,M$_{\odot}$ due to rotation induced chemically homogeneous evolution (CHE). CHE enables the merger product to enter the H-rich WNs phase at TAMS with mass greater than about 30\,M$_{\odot}$. Our results are consistent with \cite{2022A&A...667A..58P}, the H-free WNs phase appears to act as a transition phase between the H-rich WNs and WCs phase, and it typically has the shortest timescale throughout the WRs phase.

\begin{figure*}[htb]
\centering
\includegraphics[width=\textwidth]{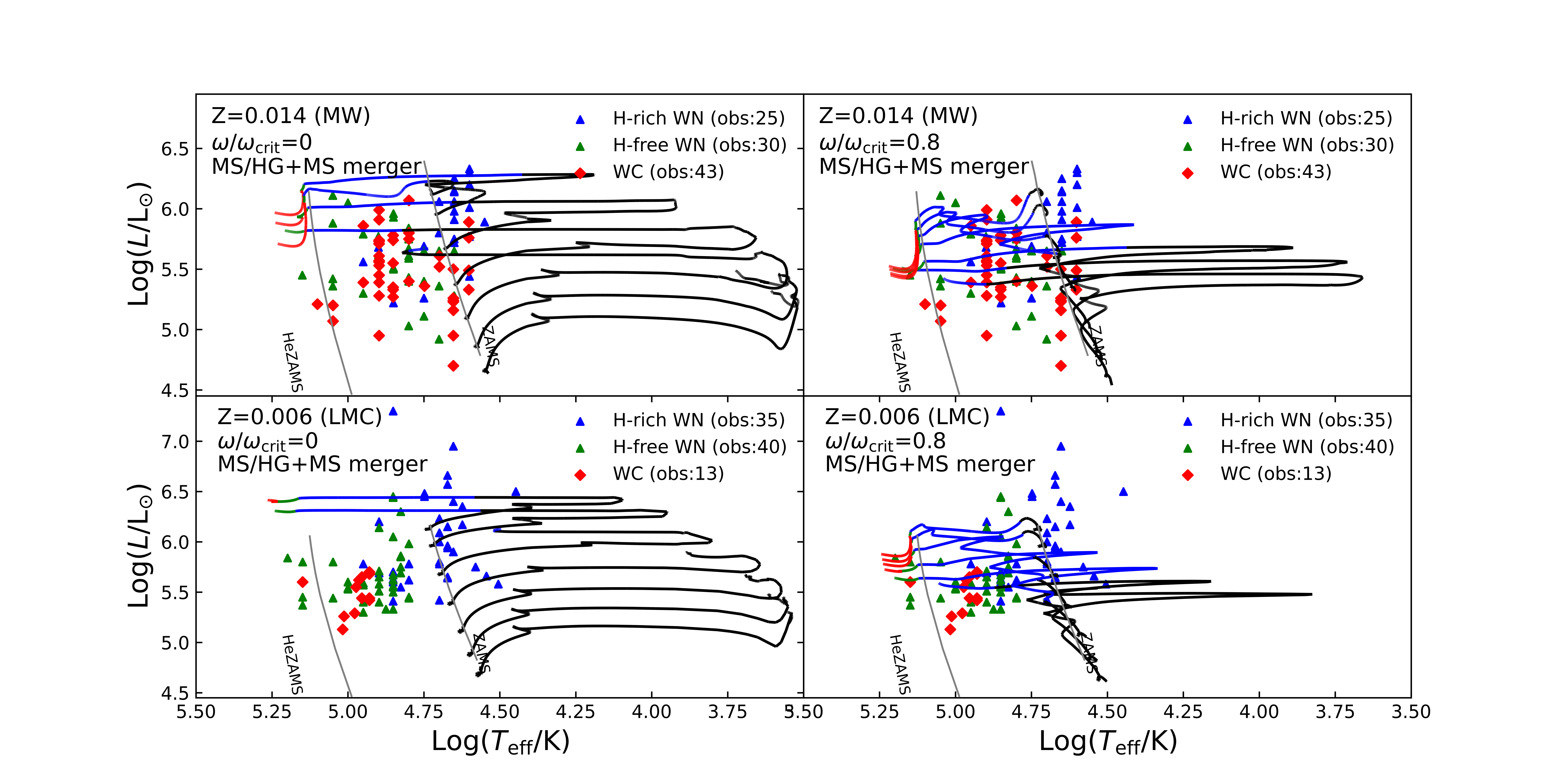}
\caption{Positions of WRs observed in the MW and LMC, and evolutionary trajectories of MS/HG + MS merger products with different masses. In the MW and LMC, the evolutionary trajectories from bottom to top correspond to initial masses of 20,24,30,40,56,80,100\,$M_{\odot}$ for the merger products. The different types/stages are marked with colours. Black: pre WRs; blue: H-rich WNs; green: H-free WNs; red: WCs. Observations of H-rich WNs and H-free WNs for MW were obtained from \cite{2019A&A...625A..57H}. Observations of WCs for MW were obtained from \cite{2019A&A...621A..92S}. Observations of H-rich WNs and H-free WNs for LMC were obtained from \cite{2014A&A...565A..27H}. Observations of WCs for LMC were obtained from \cite{2002A&A...392..653C}. It should be noted that we do not consider observations with $X_{\rm H}>0.3$.}
\label{fig:2}
\end{figure*}

Fig. \ref{fig:2} illustrates the evolutionary tracks of the merger products and compares them with the observed WRs in the HRD. The formation of larger mass products through MS/HG + MS mergers can result in stronger stellar winds, leading to the formation of WRs. When compared with observations, we find that the merger products can approximately explain only the high-luminosity WRs. For the H-rich WNs, the lower limit of the luminosity for $\omega/\omega_{\rm crit}=0$ (and $\omega/\omega_{\rm crit}=0.8$, respectively) is approximately over $10^{5.6}$\,L$_{\odot}$ ($10^{5.4}$\,L$_{\odot}$ for the MW and $10^{6.1}$\,L$_{\odot}$ ($10^{5.5}$\,L$_{\odot}$) for the LMC. For H-free WNs, only luminosities higher than about $10^{5.8}$\,L$_{\odot}$ ($10^{5.5}$\,L$_{\odot}$) for the MW and $10^{6.1}$\,L$_{\odot}$ ($10^{5.6}$\,L$_{\odot}$) for the LMC can be explained. For WCs, only luminosities higher than about $10^{5.7}$\,L$_{\odot}$ ($10^{5.4}$\,L$_{\odot}$) for the MW and $10^{6.4}$\,L$_{\odot}$ ($10^{5.7}$\,L$_{\odot}$) for the LMC. In the LMC, models without rotation have a higher lower limit for H-rich WNs luminosity than those in \cite{2022A&A...667A..58P}, which is $10^{6.0}$\,L$_{\odot}$. The main reason is that our models have a lower mass-loss rate. In addition, since the option MLT++ eliminates the effect of envelope inflation for massive stars near the Eddington limit, it leads to a much larger effective temperature for our H-free WRs model than most observed H-free WRs.

\subsubsection{Evolution of He + HG/ GB mergers}
During the He + HG/ GB merger, the ejection of the H-rich envelope outside the system suggests that these mergers \edit1{do not} necessarily require high stellar wind to become WRs. However, not all He + HG/ GB mergers can form WRs. If $\Delta E_{\rm orb}$ during the CEE is too large or too small, it can result in either successful ejection of the envelope or the retention of a significant amount of the H-rich envelope.

\begin{figure*}[htb]
\centering
\includegraphics[width=\textwidth]{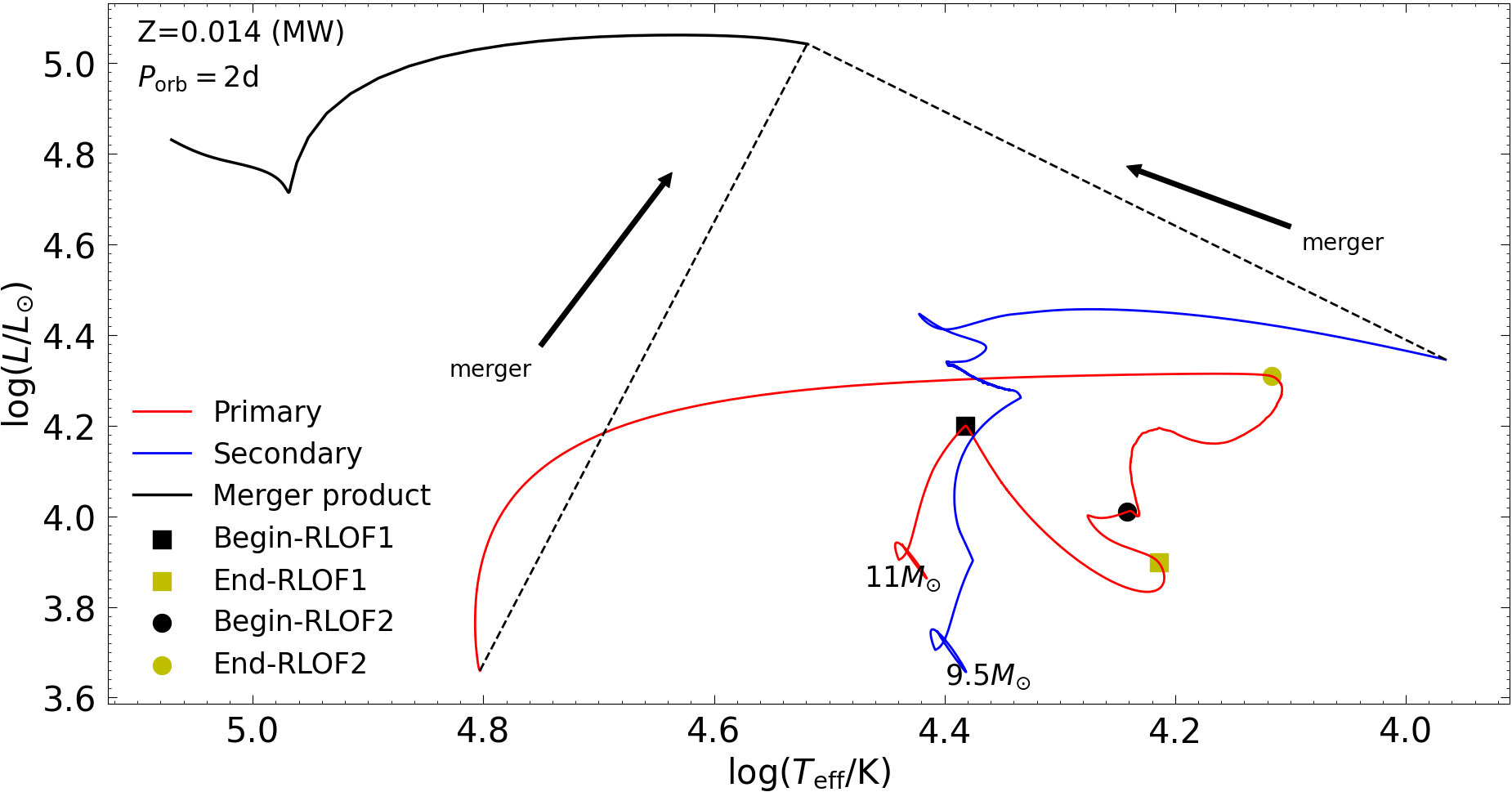}
\caption{Evolutionary tracks in HRD for WRs produced via the He + HG/GB merger channel. Initially, the components with Z = 0.014 are on the ZAMS, the primaries and the secondaries masses are 11\,M$_{\odot}$ and 9.5\,M$_{\odot}$ respectively, and the initial orbital period is 2\,d. Markers of different colors and shapes indicate the beginning and end of the Roche lobe overflow (RLOF), respectively. It is worth noting that the accretion efficiency of MT is $1-\beta=0.25$.}
\label{fig:3}
\end{figure*}

Fig. \ref{fig:3} shows a typical example of He + HG/ GB merger. Initially, the components with Z=0.014 are on the ZAMS, the primary's and the secondary's masses are 11\,M$_{\odot}$ and 9.5\,M$_{\odot}$ respectively, the initial orbital period is 2\,d and the MT is highly non-conservative, with a MT efficiency of $1-\beta=0.25$. After two stable MT, the primary evolves into a 2.3\,M$_{\odot}$ He star, and the secondary still stays in MS phase but its mass increases to 11.4\,M$_{\odot}$. Approximately 6.8\,M$_{\odot}$ of mass is lost outside the binary system (referred to as isotropic re-emission). When the secondary evolves into HG, it fills its Roche lobe and CEE occurs. At this time, $M_{1}$=2.2\,M$_{\odot}$, $M_{2}$=11.3\,M$_{\odot}$, the binary separation is 110\,R$_{\odot}$, which results in about 6.4\,M$_{\odot}$ mass of the envelope of the secondary being lost from the system during CEE. It means that the He star and its companion merger into a new star with mass of 7.1\,M$_{\odot}$.

\begin{figure*}[htb]
\centering
\includegraphics[width=\textwidth]{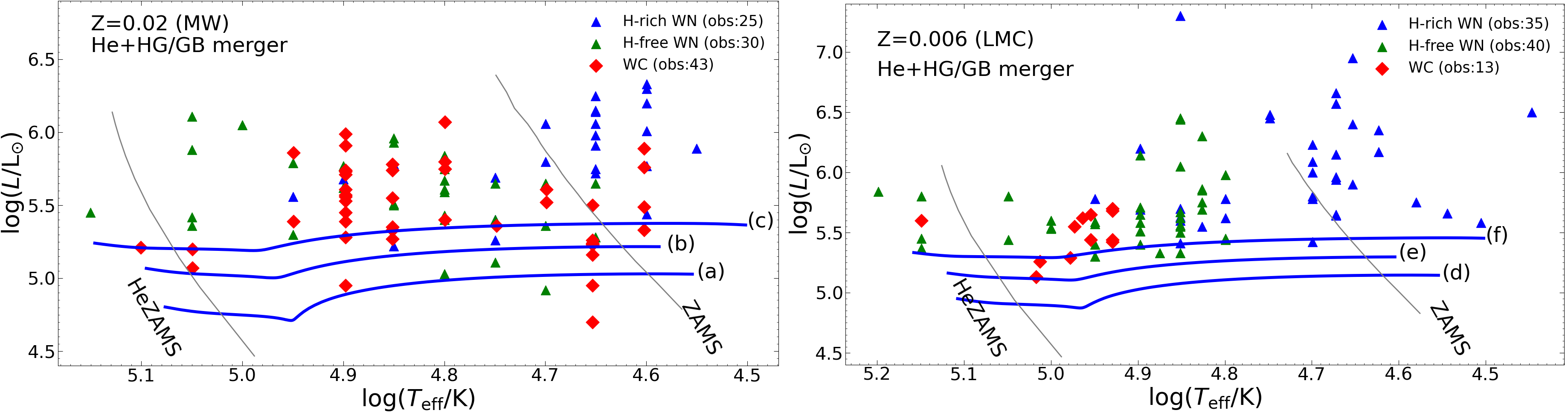}
\caption{He + HG/ GB merger product evolutionary tracks and observational samples (similar to Fig. \ref{fig:2}). For each evolutionary trajectory we describe the initial masses of the primary and secondary, the masses of the He stars and HG/ GB stars at the time of the merger, and the mass lost ($\Delta m$) of the donor due to the orbital energy changes ($\Delta E_{\rm orb}$) during CEE (i.e. $M_{\rm 1,ini} + M_{\rm 2,ini} \rightarrow M_{\rm He} + M_{\rm HG/ GB}, \Delta m$). (a): 11\,M$_{\odot}$ + 9.5\,M$_{\odot}$ $\rightarrow$ 2.3\,M$_{\odot}$ + 11.3\,M$_{\odot}$, $\Delta m$ = 6.8\,M$_{\odot}$; (b): 13\,M$_{\odot}$ + 11\,M$_{\odot}$ $\rightarrow$ 3.0\,M$_{\odot}$ + 13.1\,M$_{\odot}$, $\Delta m$ = 6.5\,M$_{\odot}$; (c): 15\,M$_{\odot}$ + 13\,M$_{\odot}$ $\rightarrow$ 3.9\,M$_{\odot}$ + 15.3\,M$_{\odot}$, $\Delta m$ = 7.0\,M$_{\odot}$; (d): 12\,M$_{\odot}$ + 10\,M$_{\odot}$ $\rightarrow$ 2.7\,M$_{\odot}$ + 12.4\,M$_{\odot}$, $\Delta m$ = 6.9\,M$_{\odot}$; (e): 14\,M$_{\odot}$ + 12\,M$_{\odot}$ $\rightarrow$ 3.4\,M$_{\odot}$ + 14.2\,M$_{\odot}$, $\Delta m$ = 6.6\,M$_{\odot}$; (f): 16\,M$_{\odot}$ + 14\,M$_{\odot}$ $\rightarrow$ 4.3\,M$_{\odot}$ + 16.4\,M$_{\odot}$, $\Delta m$ = 6.4\,M$_{\odot}$; All binary systems have an initial orbital period of 2\,days and an MT efficiency of $1-\beta = 0.25$.}
\label{fig:4}
\end{figure*}

However, the initial parameter space of the He + HG/ GB merger channel is very small, specifically, the initial primary's masses is 11\,$\sim$\,16\,M$_{\odot}$, and the initial mass ratio $q_{\rm ini}$ is $0.85\sim0.95$. This is because He stars have a very short lifetime, and most of them cannot survive until the HG/ GB stars undergo reverse MT. Thus, the mass of the merger product for this channel is roughly 7\,$\sim$\,15\,M$_{\odot}$. Fig. \ref{fig:4} shows the evolutionary tracks of the products formed via He + HG/ GB merger and compares them with the observed WRs in the HRD. After the merger, the products enter the central He-burning phase, and they evolve towards the bluer region of the HRD. Compared with observed WRs, the He + HG/ GB merger products can explain H-rich WNs with luminosities between $10^{4.7}$\,L$_{\odot}$\,$\sim$\,$10^{5.4}$\,L$_{\odot}$ for the MW and between $10^{5.0}$\,L$_{\odot}$\,$\sim$\,$10^{5.5}$\,L$_{\odot}$ for the LMC. Due to the low mass of these merger products, the stellar wind is not strong enough to completely remove the H-rich envelope, preventing these merger products from evolving into the H-free WRs phase.

\section{WR population}

Due to most WRs are affected by optically thick stellar winds, the observed effective temperature and radius of WRs are not always reliable. Therefore, comparison with the observed position of WRs in the HRD alone are often insufficient.

To address this limitation, our study utilizes population synthesis to simulate the entire population of WRs. Through this approach, we investigate both single and binary WRs systems, enabling a comprehensive analysis of their characteristics.

\subsection{Population synthesis method}

Considering that more than $50\% \sim 90\%$ of massive stars are in binary or multiple systems \citep{2012Sci...337..444S,2014ApJS..215...15S,2017ApJS..230...15M}, we assume two extreme cases for the initial binary fraction ($f_{\rm ini,bin}$) of massive stars: 50\% and 100\%. In the method of population synthesis for binary systems, we use the initial mass function of $\xi(M)\propto M^{-2.3},M\in[8,100]$\,M$_{\odot}$ for the primary components \citep{1955ApJ...121..161S,2001MNRAS.322..231K}, and the mass-ratio distribution of \edit1{$\xi(q)\propto q^{-0.1},q\in[0.1,1]\,M_{\rm 1}$} \citep{2012Sci...337..444S}. The initial orbital period distribution is $\xi(\log(P/\rm days))$ $\propto$ $\log(P/\rm days)^{-0.55}$, $\log(P/\rm days)\in[0.15,4.0]$ \citep{2012Sci...337..444S,2022A&A...667A..58P}. We assume that all initial binary systems have circular orbits. Based on the above distribution and applying Monte Carlo simulations, we create $10^{7}$ binary systems, and use MOBSE to evolve these binary systems.

In order to investigate the binary fraction of WRs, \cite{2023A&A...674A..88D} consider that a WRs is single if its radial velocity change ($\Delta RV$) satisfies $\Delta RV<50$\,km/s (the observational inclination is assumed to be $90^\circ$), or else this WRs is considered as WRs binary. In our simulations, there are some WR binaries with very wide orbits. Their $\Delta RV$ should be smaller than 50\,km/s, and these WRs can be treated as single WRs on the observations. Following \cite{2023A&A...674A..88D}, we assume that WRs are single if these WRs in binary systems at perihelion have $\Delta RV<50$\,km/s (In our grid, the orbital period is approximately greater than $10^{3.0}$$\sim$$10^{3.5}$ days), otherwise these binaries are called as WRs binaries. To distinguish them from single WRs mentioned above, WRs formed through binary mergers are referred to as merger WRs. Therefore, in this work, based on the formation scenario, isolated WRs are divided into single and merger WRs.

Using MOBSE, we evolve $10^{7}$ massive binary systems, and obtain a large grids for WRs candidates via single star and binary merger channels. Considering that MOBSE does not provide detailed stellar structure, we use MESA to evolve, cross-check, and linearly interpolate connections within the grids of different channels in the MOBSE calculations at a certain step size. For the single-star channel and the MS/ HG + MS merger channel, the mass range (initial primary or initial isolated single-star or the merger product) is roughly 20\,M$_{\odot}$\,$\sim$\,100\,M$_{\odot}$. For He + HG/ GB merger channel, the model grid covers roughly the initial primary mass in the range of 11\,M$_{\odot}$\,$\sim$\,16\,M$_{\odot}$, with the mass ratio in the range of 0.85\,$\sim$\,0.95, and the initial orbital period is in the range of 1.5\,$\sim$\,2.3\,days. In addition, for the single star channel, we assume that all models are non-rotating. For the MS/ HG  +MS merger products both $\omega=0$ and $\omega=0.8\omega_{\rm crit}$ are assumed. For the He + HG/ GB merger channel, we assume that the angular momentum is dissipated through the CEE phase, and thus the merger product of He+HG/GB is non-rotating. The above channels contain a total of 206 MESA models.

\begin{figure*}[htb]
\centering
\includegraphics[width=\textwidth]{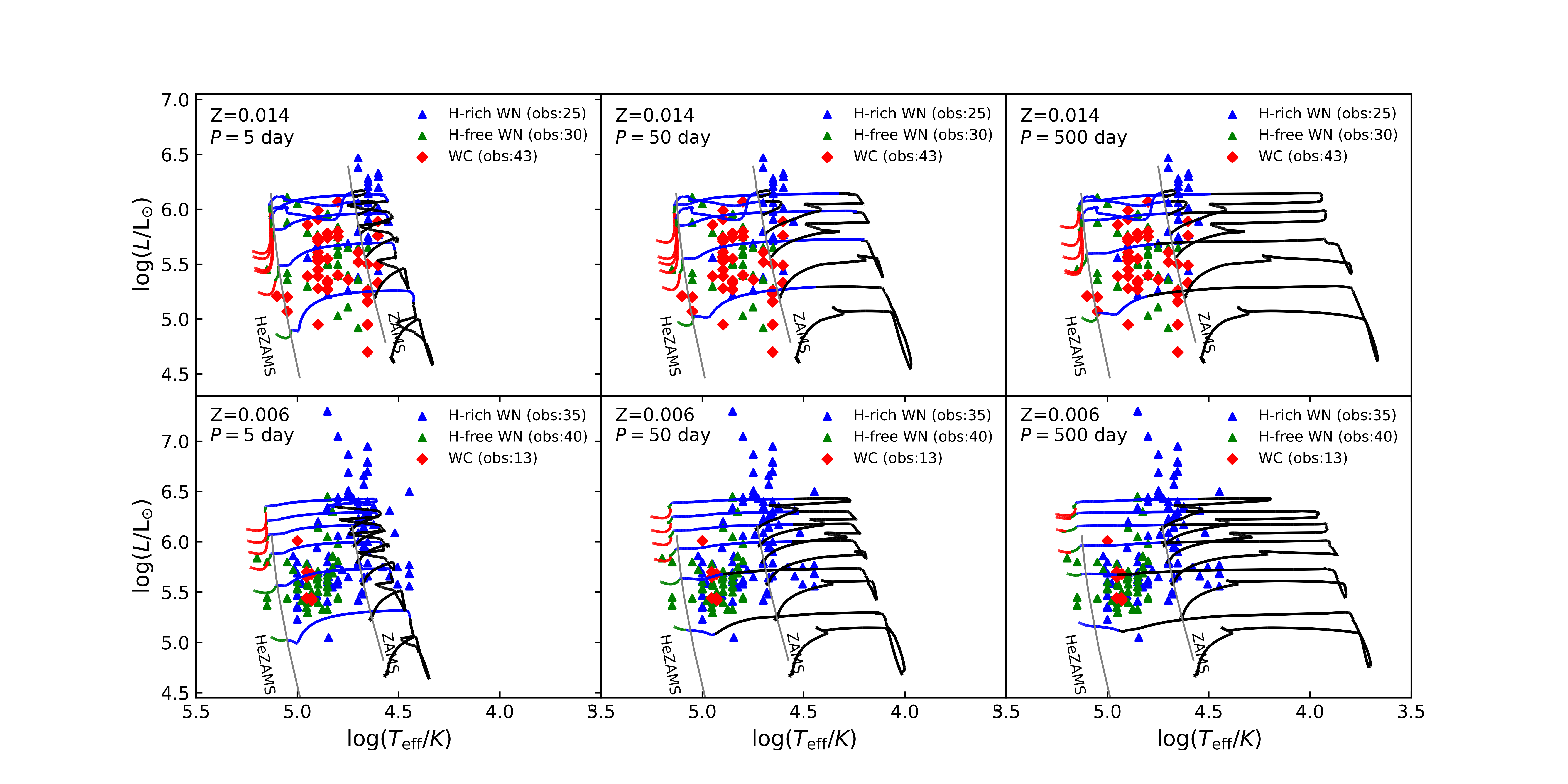}
\caption{Similar to Figs. \ref{fig:2} and \ref{fig:4}, but for WRs which are formed by the primaries through the binary channel. In all the subgraphs, the evolution trajectory from bottom to top corresponds to initial masses of the primary of 20\,M$_{\odot}$, 34\,M$_{\odot}$, 50\,M$_{\odot}$, 65\,M$_{\odot}$, 80\,M$_{\odot}$ and 100\,M$_{\odot}$ respectively. Moreover, for all these evolutionary trajectories, the initial mass ratio is consistently set at $q_{\rm ini}=0.6$.}
\label{fig:5}
\end{figure*}

Simultaneously, some massive binary systems experience MT and can eventually evolve into WRs + O-type binary systems. We also use MESA to evolve, cross-check, and linearly interpolate connections within the grids of binary non-merger channels in the MOBSE calculations at a certain step size. In MESA, we refer to \cite{2023arXiv231212658D} and treat the secondary as a point mass, mainly focusing on the structural information of the primary forming WRs. We adopt the scheme from \cite{1988A&A...202...93R} to calculate the MT. The accretion efficiency is described in Section 2.1. In this grid, the initial mass range for the primary is from 20\,M$_{\odot}$\,$\sim$\,100\,M$_{\odot}$. The range for the initial mass ratio is from 0.3\,$\sim$\,0.95. The initial orbital periods range from 5\,$\sim$\,1000 days. Through a three-dimensional interpolation of primary mass, mass ratio and orbital period, we can obtain a detail structure and evolution of primary. The binary channel grid in this work includes 2214 MESA models. Fig. \ref{fig:5} shows some of the possible results of the formation of WRs by binary channel.

\subsection{Properties of WR population}
Combining MOBSE and MESA, we obtain WR population via the method of population synthesis. This method is used in many literatures \citep{2012MNRAS.424.2265L,2017ApJ...847...62L,2019ApJ...885...20Y,2020RAA....20..161H,2021A&A...654A..57Z,2022MNRAS.515.2725G,2023RAA....23b5021Z}.

\subsubsection{WRs expected and the single/ binary fraction of WRs}
Estimating WRs' number is crucial for understanding the overall population of WRs and for conducting effective searches. Following the method in \cite{2023arXiv231212658D}, we calculate the WRs' numbers in the MW and the LMC. We assume that both the MW and LMC have a constant star formation rate (SFR). Based on \cite{2011AJ....142..197C} and \cite{2009AJ....138.1243H}, SFRs are 1.9\,M$_{\odot}$ ${\rm yr^{-1}}$ in the MW and 0.4\,M$_{\odot}$ ${\rm yr}^{-1}$ in the LMC, respectively. The WRs' number can be approximately estimated via $N_{\rm WR}=\frac{SFR \ \langle t_{\rm tot}\rangle}{\langle M \rangle} \  f_{\rm WR} \frac{\langle t_{\rm WR} \rangle}{\langle t_{\rm tot} \rangle}=\frac{SFR \ \langle t_{\rm WR}\rangle}{\langle M \rangle} \  f_{\rm WR}$, where $\langle M\rangle$ is the average mass of initial mass function; $f_{\rm WR}$ is the fraction of systems evolving into WRs; ${\langle t_{\rm tot} \rangle}$ is the average lifetime of such systems; ${\langle t_{\rm WR} \rangle}$ is the average lifetime during the WRs phase. Moreover, WRs are predominantly found as single stars in the observations, whereas most of them were expected to be in binaries according to previous theoretical models. Involving binary merger, we find that single WRs may dominate in the WR population.

\begin{figure*}[htb]
\centering
\includegraphics[width=\textwidth]{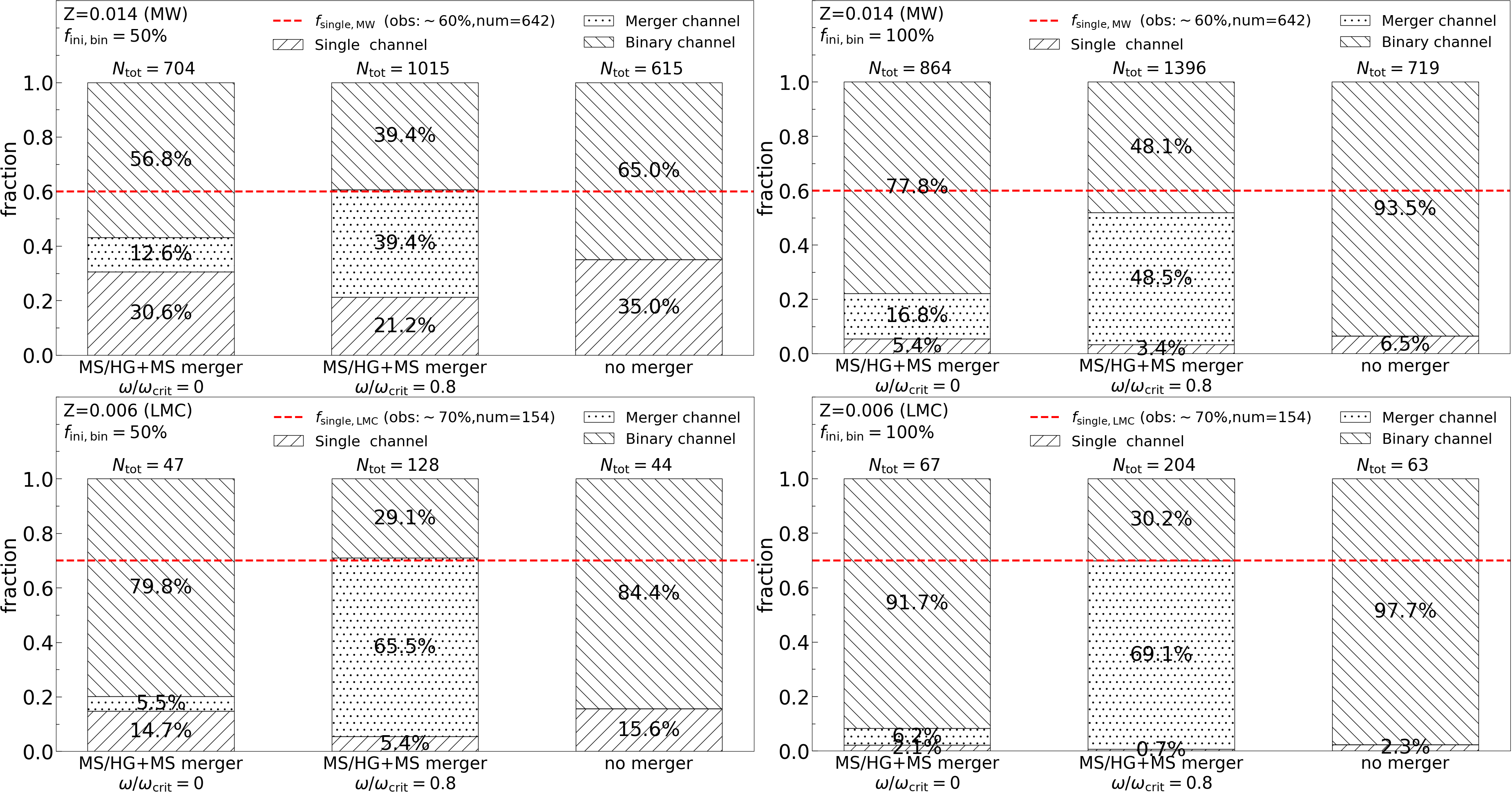}
\caption{The fractions of single and binary WRs, as well as the total number of WRs calculated in different scenarios, are determined through population synthesis. We assume different rotations for the MS/HG + MS merger product, either $\omega=0$ or $\omega=0.8\omega_{\rm crit}$, and consider initial binary fractions of $f_{\rm ini,bin}=50\%$ and 100\% for massive stars. The single channel pertains to WRs in binary systems with orbital periods exceeding approximately $10^{3.0}\sim10^{3.5}$ days ($\Delta RV<50$\,km/s) and WRs evolving from massive stars in isolated systems. The merger channel relates to single WRs resulting from binary mergers. The binary channel involves WRs binaries with orbital periods shorter than about $10^{3.0}\sim10^{3.5}$ days ($\Delta RV>50$\,km/s). These channels are represented by distinct stripes. Observations of the single and binary fractions of WRs in the MW and LMC come from \cite{2022A&A...664A..93D} and \cite{2019A&A...627A.151S}, which are shown as dashed lines. The total number of currently observed WRs in the MW and LMC is sourced from \cite{2015wrs..conf...21C}, \cite{2014ApJ...788...83M}, and \cite{2018ApJ...863..181N}.}
\label{fig:6}
\end{figure*}

Fig. \ref{fig:6} shows the calculated number of WRs and the fractions of single/ binary WRs for different cases, in which single WRs include WRs in binary systems with an orbital period longer than about 1000 days ($\Delta RV<50$\,km/s), evolved from massive stars born in isolated systems (when $f_{\rm ini,bin}=50\%$) and isolated WRs produced via binary merger. For the merger scenario our models predict 704\,$\sim$\,1396 WRs and single fraction 22.2\%\,$\sim$\,60.6\% for the MW, 47\,$\sim$\,204 WRs and single fraction 8.3\%\,$\sim$\,70.9\% for the LMC. In the MW, the number of WRs has been estimated, ranging about 6500 \citep{2001NewAR..45..135V} and 1200 \citep{1982A&A...114..409M,2015wrs..conf...21C}. Our results align with the previous estimates. So far, the known WRs in the MW and LMC are about 642 \citep{2015wrs..conf...21C} and 154 \citep{2018ApJ...863..181N}, respectively. It means that there is still a significant fraction of unknown WRs in the MW. This may be attributed to the Galactic disk and a large amount of dust obscuring many WRs. Furthermore, our simulations give upper and lower limits on the true number of WRs for the MW and LMC. Clearly, the actual number of WRs depends on $f_{\rm ini,bin}$, the rotation of MS/ HG + MS merger products, and the SFR. Considering the observation precision, the calculated single fraction of WRs is very consistent with the current observation value for the case of a high rotation rate ($\omega = 0.8\,\omega_{\rm crit}$) of the MS/HG + MS merger product.

On the other hand, the decrease in $f_{\rm ini,bin}$ and the increase in rotation of the MS/ HG + MS merger products always favours an increase in the single fraction of WRs. This is because the initial increase in the number of isolated massive stars and the rapid rotation increase the possibility of WRs formation by enhancing the mass-loss rate and extending the lifetime of WRs through enlarging the mass of the burning core \citep{2023A&A...680A.101S}. In the LMC, the very small contribution of isolated massive stars to the WR population, and the rotation induced CHE are key to the formation of WRs. Additionally, a decrease in $f_{\rm ini,bin}$ will always lead to a significant decrease in the calculated number of WRs. This is because the lifetimes of WRs formed through the single star channel are much shorter than those formed through binary channels.

Without considering mergers, the population of WRs obtained is mostly dominated by binary stars, and the calculated number of WRs in the MW is much lower than the estimates by \cite{1982A&A...114..409M}, \cite{2015wrs..conf...21C} and \cite{2001NewAR..45..135V} and also significantly lower than the observed number of WRs in the LMC. It is worth pointing out that in the LMC for $f_{\rm ini,bin}$ = 50\% our calculated binary fraction of WRs is essentially in agreement with \cite{2022A&A...667A..58P}.

\subsubsection{Luminosity distribution of WRs}
\begin{figure*}[htb]
\centering
\includegraphics[width=\textwidth]{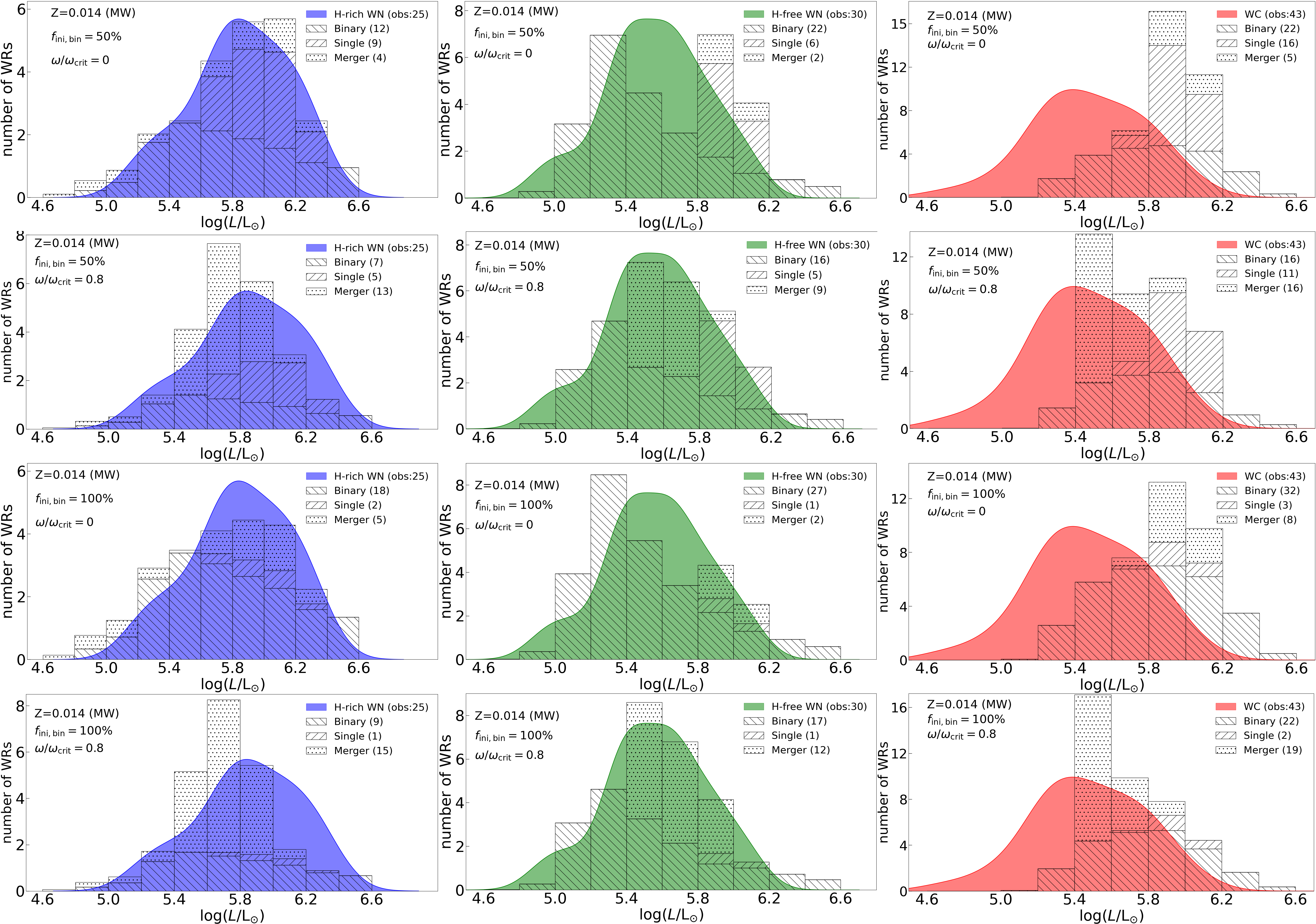}
\caption{In the MW, the luminosity distribution of different subclasses of WRs obtained after normalisation with the observed sample. Different stripes correspond to different channels. The different colours represent the different subclasses of WRs observed. Observations of H-rich WNs and H-free WNs in the MW were obtained from \cite{2019A&A...625A..57H}. Observations of WCs in the MW were obtained from \cite{2019A&A...621A..92S}. It should be noted that observations with $X_{\rm H}>0.3$ are not considered.}
\label{fig:7}
\end{figure*}

\begin{figure*}[htb]
\centering
\includegraphics[width=\textwidth]{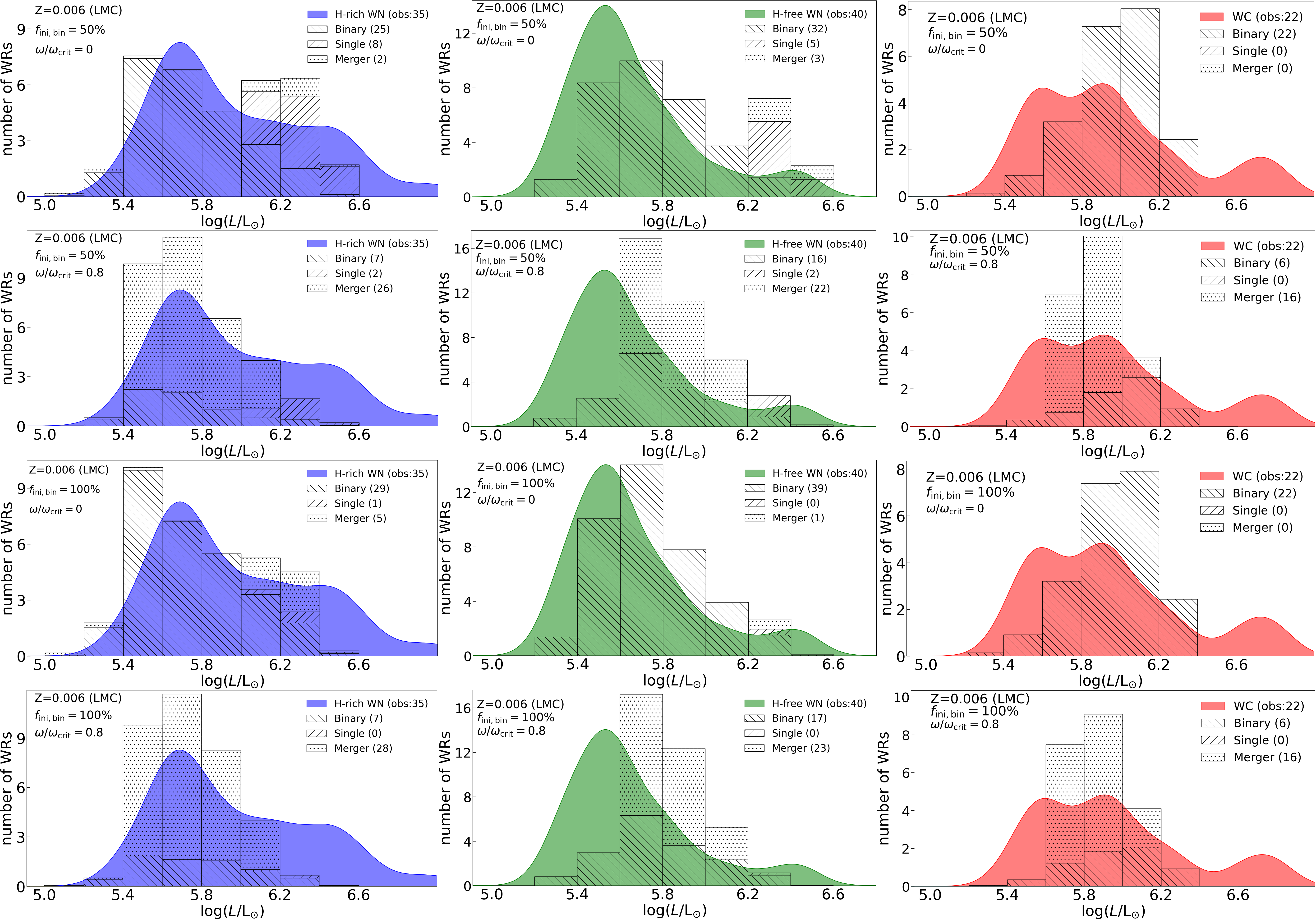}
\caption{Similar to Fig. \ref{fig:7}, but for WRs in the LMC. The luminosity distribution of different subclasses of WRs obtained after normalisation with the observed sample. Observations of H-rich WNs and H-free WNs in the LMC were obtained from \cite{2014A&A...565A..27H}. Observations of WCs in the LMC were obtained from \cite{2001MNRAS.324...18B}. Additionally, four of the 26 WCs from \cite{2001MNRAS.324...18B} were excluded based on the work of \cite{2018ApJ...863..181N}.}
\label{fig:8}
\end{figure*}

It is important to explore the luminosity distribution of WRs. Figs. \ref{fig:7} and \ref{fig:8} show the luminosity distributions of different subclasses of WRs in the MW and LMC predicted to be obtained after normalisation of the observed samples.

In Fig. \ref{fig:7}, within the MW, the luminosities of the observed WRs from different subclasses exhibit a roughly unimodal distribution. Among all subclasses, a decrease in $f_{\rm ini,bin}$ significantly increases the contribution from the single-star channel, thereby augmenting the number of WRs in the high-luminosity region ($>10^{5.8}$\,L$_{\odot}$). Rapidly rotating MS/HG + MS merger products form fainter WRs than in non-rotating scenarios due to an increased mass-loss rate, which leads to a shift of the predicted peak of the distribution toward lower luminosities. Upon comparison, we assess that the scenario where merger products are rapidly rotating roughly aligns with the observed distribution of H-rich WNs and H-free WNs. However, in the case of WCs, even with rapidly rotating MS/ HG + MS merger products, the number of low-luminosity WCs ($<10^{5.4}$\,L$_{\odot}$) in our simulation is still much lower than the observed estimate. We evaluate that appropriately enhancing the mixing efficiency of the merger products can facilitate the formation of low-luminosity WCs.

In Fig. \ref{fig:8}, the luminosities of the observed WRs from different subclasses in the LMC generally exhibit a slightly bimodal distribution. Unlike in the MW, $f_{\rm ini,bin}$ has a weaker influence on the luminosity distribution in the LMC, which is attributed to the significantly lower contribution of the single-star channel to WRs in the LMC. Among all subclasses, rapidly rotating MS/ HG + MS merger products shift the predicted peak further to lower luminosities due to CHE. Considering the observed high fraction of single WRs in the LMC, we only compare scenarios where MS/HG + MS merger products are rapidly rotating. For H-rich WNs, H-free WNs, and WCs, the luminosity distributions simulated in our models roughly match the first peak of the observed distributions. However, the second peak observed in the samples ($\sim10^{6.4}$\,L$_{\odot}$ for H-rich WNs and H-free WNs, and $\sim10^{6.8}$\,L$_{\odot}$ for WCs) does not appear in our simulations. We evaluate that appropriately increasing the mass-loss rate in the single-star channel could potentially improve this discrepancy, as the single-star channel's contribution to WRs luminosities primarily fall between $10^{6.0}$\,L$_{\odot}$ and $10^{6.6}$\,L$_{\odot}$.

\section{CONCLUSIONS}

Considering binary interaction and using MOBSE and MESA code, we investigate the WRs formation via massive star merger.
We find that the massive-star merger can successfully produce WRs. The MS/ HG + MS merger, producing a massive star with high rotation and high mass-loss rate, can evolve into the WRs with luminosity higher than $10^{5.4}$\,L$_{\odot}$. The He + HG/ GB merger, ejecting most of the H-rich envelopes, can produce the WRs with low luminosity WRs of $10^{4.7}$\,L$_{\odot}$\,$\sim$\,$10^{5.5}$\,L$_{\odot}$.

Combining MOBSE and MESA codes and using the method of population synthesis, we simulate WR population. We estimate that there are approximately 704$\sim$1396 WRs in the MW and 47$\,\sim$\,204 WRs in the LMC.
Here, the contribution of WRs produced via binary merger to total WRs is about 12.6\%\,$\sim$\,48.5\% for the MW and 5.5\%\,$\sim$\,69.1\% for the LMC. In addition, the fraction of single WRs in whole WRs is about 51.9\%\,$\sim$\,60.6\% for the MW and 69.8\%\,$\sim$\,70.9\% for the LMC in the models with rapid rotation after binary merger, which is consistent with the observed estimate. These indicate that the binary merger is significant formation scenario for WRs formation, and can explain the high fraction of the single WRs in the total population.

\section*{Acknowledgements}
This work received the support of the National Natural Science Foundation of China under grants U2031204, 12373038, 12163005, and 12288102, the science research grants from the China Manned Space Project with No. CMSCSST-2021-A10, and the Natural Science Foundation of Xinjiang Nos. 2021D01C075, No.2022D01D85, and 2022TSYCLJ0006.

\bibliography{./sample631}

\begin{thebibliography}{}
\expandafter\ifx\csname natexlab\endcsname\relax\def\natexlab#1{#1}\fi
\providecommand{\url}[1]{\href{#1}{#1}}
\providecommand{\dodoi}[1]{doi:~\href{http://doi.org/#1}{\nolinkurl{#1}}}
\providecommand{\doeprint}[1]{\href{http://ascl.net/#1}{\nolinkurl{http://ascl.net/#1}}}
\providecommand{\doarXiv}[1]{\href{https://arxiv.org/abs/#1}{\nolinkurl{https://arxiv.org/abs/#1}}}

\bibitem[{{Aguilera-Dena} {et~al.}(2022){Aguilera-Dena}, {Langer},
  {Antoniadis}, {Pauli}, {Dessart}, {Vigna-G{\'o}mez}, {Gr{\"a}fener}, \&
  {Yoon}}]{2022A&A...661A..60A}
{Aguilera-Dena}, D.~R., {Langer}, N., {Antoniadis}, J., {et~al.} 2022, \aap,
  661, A60, \dodoi{10.1051/0004-6361/202142895}

\bibitem[{{Bartzakos} {et~al.}(2001){Bartzakos}, {Moffat}, \&
  {Niemela}}]{2001MNRAS.324...18B}
{Bartzakos}, P., {Moffat}, A.~F.~J., \& {Niemela}, V.~S. 2001, \mnras, 324, 18,
  \dodoi{10.1046/j.1365-8711.2001.04126.x}

\bibitem[{{Beasor} {et~al.}(2020){Beasor}, {Davies}, {Smith}, {van Loon},
  {Gehrz}, \& {Figer}}]{2020MNRAS.492.5994B}
{Beasor}, E.~R., {Davies}, B., {Smith}, N., {et~al.} 2020, \mnras, 492, 5994,
  \dodoi{10.1093/mnras/staa255}

\bibitem[{{Bj{\"o}rklund} {et~al.}(2021){Bj{\"o}rklund}, {Sundqvist}, {Puls},
  \& {Najarro}}]{2021A&A...648A..36B}
{Bj{\"o}rklund}, R., {Sundqvist}, J.~O., {Puls}, J., \& {Najarro}, F. 2021,
  \aap, 648, A36, \dodoi{10.1051/0004-6361/202038384}

\bibitem[{{Brott} {et~al.}(2011){Brott}, {de Mink}, {Cantiello}, {Langer}, {de
  Koter}, {Evans}, {Hunter}, {Trundle}, \& {Vink}}]{2011A&A...530A.115B}
{Brott}, I., {de Mink}, S.~E., {Cantiello}, M., {et~al.} 2011, \aap, 530, A115,
  \dodoi{10.1051/0004-6361/201016113}

\bibitem[{{Chomiuk} \& {Povich}(2011)}]{2011AJ....142..197C}
{Chomiuk}, L., \& {Povich}, M.~S. 2011, \aj, 142, 197,
  \dodoi{10.1088/0004-6256/142/6/197}

\bibitem[{{Costa} {et~al.}(2022){Costa}, {Ballone}, {Mapelli}, \&
  {Bressan}}]{2022MNRAS.516.1072C}
{Costa}, G., {Ballone}, A., {Mapelli}, M., \& {Bressan}, A. 2022, \mnras, 516,
  1072, \dodoi{10.1093/mnras/stac2222}

\bibitem[{{Crowther}(2007)}]{2007ARA&A..45..177C}
{Crowther}, P.~A. 2007, \araa, 45, 177,
  \dodoi{10.1146/annurev.astro.45.051806.110615}

\bibitem[{{Crowther}(2015)}]{2015wrs..conf...21C}
{Crowther}, P.~A. 2015, in Wolf-Rayet Stars, ed. W.-R. {Hamann}, A.~{Sander},
  \& H.~{Todt}, 21--26, \dodoi{10.48550/arXiv.1509.00495}

\bibitem[{{Crowther} {et~al.}(2002){Crowther}, {Dessart}, {Hillier}, {Abbott},
  \& {Fullerton}}]{2002A&A...392..653C}
{Crowther}, P.~A., {Dessart}, L., {Hillier}, D.~J., {Abbott}, J.~B., \&
  {Fullerton}, A.~W. 2002, \aap, 392, 653, \dodoi{10.1051/0004-6361:20020941}

\bibitem[{{Cui} {et~al.}(2018){Cui}, {Wang}, {Zhu}, {L{\"u}}, {Chen}, \&
  {Han}}]{2018PASP..130h4202C}
{Cui}, Z., {Wang}, Z., {Zhu}, C., {et~al.} 2018, \pasp, 130, 084202,
  \dodoi{10.1088/1538-3873/aac55e}

\bibitem[{{de Mink} {et~al.}(2014){de Mink}, {Sana}, {Langer}, {Izzard}, \&
  {Schneider}}]{2014ApJ...782....7D}
{de Mink}, S.~E., {Sana}, H., {Langer}, N., {Izzard}, R.~G., \& {Schneider},
  F.~R.~N. 2014, \apj, 782, 7, \dodoi{10.1088/0004-637X/782/1/7}

\bibitem[{{Dominik} {et~al.}(2012){Dominik}, {Belczynski}, {Fryer}, {Holz},
  {Berti}, {Bulik}, {Mandel}, \& {O'Shaughnessy}}]{2012ApJ...759...52D}
{Dominik}, M., {Belczynski}, K., {Fryer}, C., {et~al.} 2012, \apj, 759, 52,
  \dodoi{10.1088/0004-637X/759/1/52}

\bibitem[{{Dsilva} {et~al.}(2020){Dsilva}, {Shenar}, {Sana}, \&
  {Marchant}}]{2020A&A...641A..26D}
{Dsilva}, K., {Shenar}, T., {Sana}, H., \& {Marchant}, P. 2020, \aap, 641, A26,
  \dodoi{10.1051/0004-6361/202038446}

\bibitem[{{Dsilva} {et~al.}(2022){Dsilva}, {Shenar}, {Sana}, \&
  {Marchant}}]{2022A&A...664A..93D}
---. 2022, \aap, 664, A93, \dodoi{10.1051/0004-6361/202142729}

\bibitem[{{Dsilva} {et~al.}(2023){Dsilva}, {Shenar}, {Sana}, \&
  {Marchant}}]{2023A&A...674A..88D}
---. 2023, \aap, 674, A88, \dodoi{10.1051/0004-6361/202244308}

\bibitem[{{Dutta} \& {Klencki}(2023)}]{2023arXiv231212658D}
{Dutta}, D., \& {Klencki}, J. 2023, arXiv e-prints, arXiv:2312.12658,
  \dodoi{10.48550/arXiv.2312.12658}

\bibitem[{{Eggenberger} {et~al.}(2021){Eggenberger}, {Ekstr{\"o}m}, {Georgy},
  {Martinet}, {Pezzotti}, {Nandal}, {Meynet}, {Buldgen}, {Salmon},
  {Haemmerl{\'e}}, {Maeder}, {Hirschi}, {Yusof}, {Groh}, {Farrell}, {Murphy},
  \& {Choplin}}]{2021A&A...652A.137E}
{Eggenberger}, P., {Ekstr{\"o}m}, S., {Georgy}, C., {et~al.} 2021, \aap, 652,
  A137, \dodoi{10.1051/0004-6361/202141222}

\bibitem[{{Eggleton}(1983)}]{1983ApJ...268..368E}
{Eggleton}, P.~P. 1983, \apj, 268, 368, \dodoi{10.1086/160960}

\bibitem[{{Ekstr{\"o}m} {et~al.}(2012){Ekstr{\"o}m}, {Georgy}, {Eggenberger},
  {Meynet}, {Mowlavi}, {Wyttenbach}, {Granada}, {Decressin}, {Hirschi},
  {Frischknecht}, {Charbonnel}, \& {Maeder}}]{2012A&A...537A.146E}
{Ekstr{\"o}m}, S., {Georgy}, C., {Eggenberger}, P., {et~al.} 2012, \aap, 537,
  A146, \dodoi{10.1051/0004-6361/201117751}

\bibitem[{{Eldridge} {et~al.}(2017){Eldridge}, {Stanway}, {Xiao}, {McClelland},
  {Taylor}, {Ng}, {Greis}, \& {Bray}}]{2017PASA...34...58E}
{Eldridge}, J.~J., {Stanway}, E.~R., {Xiao}, L., {et~al.} 2017, \pasa, 34,
  e058, \dodoi{10.1017/pasa.2017.51}

\bibitem[{{Farrell} {et~al.}(2020){Farrell}, {Groh}, {Meynet}, {Eldridge},
  {Ekstr{\"o}m}, \& {Georgy}}]{2020MNRAS.495.4659F}
{Farrell}, E.~J., {Groh}, J.~H., {Meynet}, G., {et~al.} 2020, \mnras, 495,
  4659, \dodoi{10.1093/mnras/staa1360}

\bibitem[{{Foellmi} {et~al.}(2003{\natexlab{a}}){Foellmi}, {Moffat}, \&
  {Guerrero}}]{2003MNRAS.338..360F}
{Foellmi}, C., {Moffat}, A.~F.~J., \& {Guerrero}, M.~A. 2003{\natexlab{a}},
  \mnras, 338, 360, \dodoi{10.1046/j.1365-8711.2003.06052.x}

\bibitem[{{Foellmi} {et~al.}(2003{\natexlab{b}}){Foellmi}, {Moffat}, \&
  {Guerrero}}]{2003MNRAS.338.1025F}
---. 2003{\natexlab{b}}, \mnras, 338, 1025,
  \dodoi{10.1046/j.1365-8711.2003.06161.x}

\bibitem[{{Fryer} \& {Heger}(2005)}]{2005ApJ...623..302F}
{Fryer}, C.~L., \& {Heger}, A. 2005, \apj, 623, 302, \dodoi{10.1086/428379}

\bibitem[{{Fullerton} {et~al.}(2006){Fullerton}, {Massa}, \&
  {Prinja}}]{2006ApJ...637.1025F}
{Fullerton}, A.~W., {Massa}, D.~L., \& {Prinja}, R.~K. 2006, \apj, 637, 1025,
  \dodoi{10.1086/498560}

\bibitem[{{Gaburov} {et~al.}(2008){Gaburov}, {Lombardi}, \& {Portegies
  Zwart}}]{2008MNRAS.383L...5G}
{Gaburov}, E., {Lombardi}, J.~C., \& {Portegies Zwart}, S. 2008, \mnras, 383,
  L5, \dodoi{10.1111/j.1745-3933.2007.00399.x}

\bibitem[{{Georgy} {et~al.}(2012){Georgy}, {Ekstr{\"o}m}, {Meynet}, {Massey},
  {Levesque}, {Hirschi}, {Eggenberger}, \& {Maeder}}]{2012A&A...542A..29G}
{Georgy}, C., {Ekstr{\"o}m}, S., {Meynet}, G., {et~al.} 2012, \aap, 542, A29,
  \dodoi{10.1051/0004-6361/201118340}

\bibitem[{{Giacobbo} {et~al.}(2018){Giacobbo}, {Mapelli}, \&
  {Spera}}]{2018MNRAS.474.2959G}
{Giacobbo}, N., {Mapelli}, M., \& {Spera}, M. 2018, \mnras, 474, 2959,
  \dodoi{10.1093/mnras/stx2933}

\bibitem[{{Glebbeek} {et~al.}(2013){Glebbeek}, {Gaburov}, {Portegies Zwart}, \&
  {Pols}}]{2013MNRAS.434.3497G}
{Glebbeek}, E., {Gaburov}, E., {Portegies Zwart}, S., \& {Pols}, O.~R. 2013,
  \mnras, 434, 3497, \dodoi{10.1093/mnras/stt1268}

\bibitem[{{Gr{\"a}fener} {et~al.}(2012){Gr{\"a}fener}, {Owocki}, \&
  {Vink}}]{2012A&A...538A..40G}
{Gr{\"a}fener}, G., {Owocki}, S.~P., \& {Vink}, J.~S. 2012, \aap, 538, A40,
  \dodoi{10.1051/0004-6361/201117497}

\bibitem[{{Guo} {et~al.}(2022){Guo}, {Wang}, \& {Han}}]{2022MNRAS.515.2725G}
{Guo}, Y., {Wang}, B., \& {Han}, Z. 2022, \mnras, 515, 2725,
  \dodoi{10.1093/mnras/stac1917}

\bibitem[{{Hainich} {et~al.}(2014){Hainich}, {R{\"u}hling}, {Todt}, {Oskinova},
  {Liermann}, {Gr{\"a}fener}, {Foellmi}, {Schnurr}, \&
  {Hamann}}]{2014A&A...565A..27H}
{Hainich}, R., {R{\"u}hling}, U., {Todt}, H., {et~al.} 2014, \aap, 565, A27,
  \dodoi{10.1051/0004-6361/201322696}

\bibitem[{{Hamann} {et~al.}(2019){Hamann}, {Gr{\"a}fener}, {Liermann},
  {Hainich}, {Sander}, {Shenar}, {Ramachandran}, {Todt}, \&
  {Oskinova}}]{2019A&A...625A..57H}
{Hamann}, W.~R., {Gr{\"a}fener}, G., {Liermann}, A., {et~al.} 2019, \aap, 625,
  A57, \dodoi{10.1051/0004-6361/201834850}

\bibitem[{{Han} {et~al.}(1995){Han}, {Podsiadlowski}, \&
  {Eggleton}}]{1995MNRAS.272..800H}
{Han}, Z., {Podsiadlowski}, P., \& {Eggleton}, P.~P. 1995, \mnras, 272, 800,
  \dodoi{10.1093/mnras/272.4.800}

\bibitem[{{Han} {et~al.}(2020){Han}, {Ge}, {Chen}, \&
  {Chen}}]{2020RAA....20..161H}
{Han}, Z.-W., {Ge}, H.-W., {Chen}, X.-F., \& {Chen}, H.-L. 2020, Research in
  Astronomy and Astrophysics, 20, 161, \dodoi{10.1088/1674-4527/20/10/161}

\bibitem[{{Harris} \& {Zaritsky}(2009)}]{2009AJ....138.1243H}
{Harris}, J., \& {Zaritsky}, D. 2009, \aj, 138, 1243,
  \dodoi{10.1088/0004-6256/138/5/1243}

\bibitem[{{He} {et~al.}(2024){He}, {Shao}, {Xu}, \& {Li}}]{2024MNRAS.529.1886H}
{He}, J.-G., {Shao}, Y., {Xu}, X.-J., \& {Li}, X.-D. 2024, \mnras, 529, 1886,
  \dodoi{10.1093/mnras/stae683}

\bibitem[{{Heger} {et~al.}(2003){Heger}, {Fryer}, {Woosley}, {Langer}, \&
  {Hartmann}}]{2003ApJ...591..288H}
{Heger}, A., {Fryer}, C.~L., {Woosley}, S.~E., {Langer}, N., \& {Hartmann},
  D.~H. 2003, \apj, 591, 288, \dodoi{10.1086/375341}

\bibitem[{{Heger} {et~al.}(2000){Heger}, {Langer}, \&
  {Woosley}}]{2000ApJ...528..368H}
{Heger}, A., {Langer}, N., \& {Woosley}, S.~E. 2000, \apj, 528, 368,
  \dodoi{10.1086/308158}

\bibitem[{{Hellings}(1983)}]{1983Ap&SS..96...37H}
{Hellings}, P. 1983, \apss, 96, 37, \dodoi{10.1007/BF00661941}

\bibitem[{{Hjellming} \& {Webbink}(1987)}]{1987ApJ...318..794H}
{Hjellming}, M.~S., \& {Webbink}, R.~F. 1987, \apj, 318, 794,
  \dodoi{10.1086/165412}

\bibitem[{{Hurley} {et~al.}(2000){Hurley}, {Pols}, \&
  {Tout}}]{2000MNRAS.315..543H}
{Hurley}, J.~R., {Pols}, O.~R., \& {Tout}, C.~A. 2000, \mnras, 315, 543,
  \dodoi{10.1046/j.1365-8711.2000.03426.x}

\bibitem[{{Hurley} {et~al.}(2002){Hurley}, {Tout}, \&
  {Pols}}]{2002MNRAS.329..897H}
{Hurley}, J.~R., {Tout}, C.~A., \& {Pols}, O.~R. 2002, \mnras, 329, 897,
  \dodoi{10.1046/j.1365-8711.2002.05038.x}

\bibitem[{{Ivanova} \& {Taam}(2004)}]{2004ApJ...601.1058I}
{Ivanova}, N., \& {Taam}, R.~E. 2004, \apj, 601, 1058, \dodoi{10.1086/380561}

\bibitem[{{Ivanova} {et~al.}(2013){Ivanova}, {Justham}, {Chen}, {De Marco},
  {Fryer}, {Gaburov}, {Ge}, {Glebbeek}, {Han}, {Li}, {Lu}, {Marsh},
  {Podsiadlowski}, {Potter}, {Soker}, {Taam}, {Tauris}, {van den Heuvel}, \&
  {Webbink}}]{2013A&ARv..21...59I}
{Ivanova}, N., {Justham}, S., {Chen}, X., {et~al.} 2013, \aapr, 21, 59,
  \dodoi{10.1007/s00159-013-0059-2}

\bibitem[{{Kiel} \& {Hurley}(2006)}]{2006MNRAS.369.1152K}
{Kiel}, P.~D., \& {Hurley}, J.~R. 2006, \mnras, 369, 1152,
  \dodoi{10.1111/j.1365-2966.2006.10400.x}

\bibitem[{{Kippenhahn} {et~al.}(1980){Kippenhahn}, {Ruschenplatt}, \&
  {Thomas}}]{1980A&A....91..175K}
{Kippenhahn}, R., {Ruschenplatt}, G., \& {Thomas}, H.~C. 1980, \aap, 91, 175

\bibitem[{{Kroupa}(2001)}]{2001MNRAS.322..231K}
{Kroupa}, P. 2001, \mnras, 322, 231, \dodoi{10.1046/j.1365-8711.2001.04022.x}

\bibitem[{{Krti{\v{c}}ka} \& {Kub{\'a}t}(2017)}]{2017A&A...606A..31K}
{Krti{\v{c}}ka}, J., \& {Kub{\'a}t}, J. 2017, \aap, 606, A31,
  \dodoi{10.1051/0004-6361/201730723}

\bibitem[{{Langer}(1991)}]{1991A&A...252..669L}
{Langer}, N. 1991, \aap, 252, 669

\bibitem[{{Langer}(1998)}]{1998A&A...329..551L}
---. 1998, \aap, 329, 551

\bibitem[{{Langer}(2012)}]{2012ARA&A..50..107L}
---. 2012, \araa, 50, 107, \dodoi{10.1146/annurev-astro-081811-125534}

\bibitem[{{Langer} {et~al.}(1994){Langer}, {Hamann}, {Lennon}, {Najarro},
  {Pauldrach}, \& {Puls}}]{1994A&A...290..819L}
{Langer}, N., {Hamann}, W.~R., {Lennon}, M., {et~al.} 1994, \aap, 290, 819

\bibitem[{{Lombardi} {et~al.}(2002){Lombardi}, {Warren}, {Rasio}, {Sills}, \&
  {Warren}}]{2002ApJ...568..939L}
{Lombardi}, James~C., J., {Warren}, J.~S., {Rasio}, F.~A., {Sills}, A., \&
  {Warren}, A.~R. 2002, \apj, 568, 939, \dodoi{10.1086/339060}

\bibitem[{{L{\"u}} {et~al.}(2017){L{\"u}}, {Zhu}, {Wang}, \&
  {Iminniyaz}}]{2017ApJ...847...62L}
{L{\"u}}, G., {Zhu}, C., {Wang}, Z., \& {Iminniyaz}, H. 2017, \apj, 847, 62,
  \dodoi{10.3847/1538-4357/aa8a77}

\bibitem[{{L{\"u}} {et~al.}(2012){L{\"u}}, {Zhu}, {Postnov}, {Yungelson},
  {Kuranov}, \& {Wang}}]{2012MNRAS.424.2265L}
{L{\"u}}, G.~L., {Zhu}, C.~H., {Postnov}, K.~A., {et~al.} 2012, \mnras, 424,
  2265, \dodoi{10.1111/j.1365-2966.2012.21395.x}

\bibitem[{{Lu} {et~al.}(2023){Lu}, {Zhu}, {Liu}, {Guo}, {Yu}, \&
  {L{\"u}}}]{2023A&A...674A.216L}
{Lu}, X., {Zhu}, C., {Liu}, H., {et~al.} 2023, \aap, 674, A216,
  \dodoi{10.1051/0004-6361/202243188}

\bibitem[{{Maeder} \& {Lequeux}(1982)}]{1982A&A...114..409M}
{Maeder}, A., \& {Lequeux}, J. 1982, \aap, 114, 409

\bibitem[{{Massey} {et~al.}(2014){Massey}, {Neugent}, {Morrell}, \&
  {Hillier}}]{2014ApJ...788...83M}
{Massey}, P., {Neugent}, K.~F., {Morrell}, N., \& {Hillier}, D.~J. 2014, \apj,
  788, 83, \dodoi{10.1088/0004-637X/788/1/83}

\bibitem[{{Mauron} \& {Josselin}(2011)}]{2011A&A...526A.156M}
{Mauron}, N., \& {Josselin}, E. 2011, \aap, 526, A156,
  \dodoi{10.1051/0004-6361/201013993}

\bibitem[{{Menon} {et~al.}(2021){Menon}, {Langer}, {de Mink}, {Justham}, {Sen},
  {Sz{\'e}csi}, {de Koter}, {Abdul-Masih}, {Sana}, {Mahy}, \&
  {Marchant}}]{2021MNRAS.507.5013M}
{Menon}, A., {Langer}, N., {de Mink}, S.~E., {et~al.} 2021, \mnras, 507, 5013,
  \dodoi{10.1093/mnras/stab2276}

\bibitem[{{Meynet} \& {Maeder}(2003)}]{2003A&A...404..975M}
{Meynet}, G., \& {Maeder}, A. 2003, \aap, 404, 975,
  \dodoi{10.1051/0004-6361:20030512}

\bibitem[{{Meynet} \& {Maeder}(2005)}]{2005A&A...429..581M}
---. 2005, \aap, 429, 581, \dodoi{10.1051/0004-6361:20047106}

\bibitem[{{Moe} \& {Di Stefano}(2017)}]{2017ApJS..230...15M}
{Moe}, M., \& {Di Stefano}, R. 2017, \apjs, 230, 15,
  \dodoi{10.3847/1538-4365/aa6fb6}

\bibitem[{{Neugent} \& {Massey}(2014)}]{2014ApJ...789...10N}
{Neugent}, K.~F., \& {Massey}, P. 2014, \apj, 789, 10,
  \dodoi{10.1088/0004-637X/789/1/10}

\bibitem[{{Neugent} {et~al.}(2018){Neugent}, {Massey}, \&
  {Morrell}}]{2018ApJ...863..181N}
{Neugent}, K.~F., {Massey}, P., \& {Morrell}, N. 2018, \apj, 863, 181,
  \dodoi{10.3847/1538-4357/aad17d}

\bibitem[{{Nieuwenhuijzen} \& {de Jager}(1990)}]{1990A&A...231..134N}
{Nieuwenhuijzen}, H., \& {de Jager}, C. 1990, \aap, 231, 134

\bibitem[{{Nugis} \& {Lamers}(2000)}]{2000A&A...360..227N}
{Nugis}, T., \& {Lamers}, H.~J.~G.~L.~M. 2000, \aap, 360, 227

\bibitem[{{Pauli} {et~al.}(2022){Pauli}, {Langer}, {Aguilera-Dena}, {Wang}, \&
  {Marchant}}]{2022A&A...667A..58P}
{Pauli}, D., {Langer}, N., {Aguilera-Dena}, D.~R., {Wang}, C., \& {Marchant},
  P. 2022, \aap, 667, A58, \dodoi{10.1051/0004-6361/202243965}

\bibitem[{{Paxton} {et~al.}(2011){Paxton}, {Bildsten}, {Dotter}, {Herwig},
  {Lesaffre}, \& {Timmes}}]{2011ApJS..192....3P}
{Paxton}, B., {Bildsten}, L., {Dotter}, A., {et~al.} 2011, \apjs, 192, 3,
  \dodoi{10.1088/0067-0049/192/1/3}

\bibitem[{{Paxton} {et~al.}(2013){Paxton}, {Cantiello}, {Arras}, {Bildsten},
  {Brown}, {Dotter}, {Mankovich}, {Montgomery}, {Stello}, {Timmes}, \&
  {Townsend}}]{2013ApJS..208....4P}
{Paxton}, B., {Cantiello}, M., {Arras}, P., {et~al.} 2013, \apjs, 208, 4,
  \dodoi{10.1088/0067-0049/208/1/4}

\bibitem[{{Paxton} {et~al.}(2015){Paxton}, {Marchant}, {Schwab}, {Bauer},
  {Bildsten}, {Cantiello}, {Dessart}, {Farmer}, {Hu}, {Langer}, {Townsend},
  {Townsley}, \& {Timmes}}]{2015ApJS..220...15P}
{Paxton}, B., {Marchant}, P., {Schwab}, J., {et~al.} 2015, \apjs, 220, 15,
  \dodoi{10.1088/0067-0049/220/1/15}

\bibitem[{{Paxton} {et~al.}(2018){Paxton}, {Schwab}, {Bauer}, {Bildsten},
  {Blinnikov}, {Duffell}, {Farmer}, {Goldberg}, {Marchant}, {Sorokina},
  {Thoul}, {Townsend}, \& {Timmes}}]{2018ApJS..234...34P}
{Paxton}, B., {Schwab}, J., {Bauer}, E.~B., {et~al.} 2018, \apjs, 234, 34,
  \dodoi{10.3847/1538-4365/aaa5a8}

\bibitem[{{Podsiadlowski} {et~al.}(1992){Podsiadlowski}, {Joss}, \&
  {Hsu}}]{1992ApJ...391..246P}
{Podsiadlowski}, P., {Joss}, P.~C., \& {Hsu}, J.~J.~L. 1992, \apj, 391, 246,
  \dodoi{10.1086/171341}

\bibitem[{{Pols} {et~al.}(1998){Pols}, {Schr{\"o}der}, {Hurley}, {Tout}, \&
  {Eggleton}}]{1998MNRAS.298..525P}
{Pols}, O.~R., {Schr{\"o}der}, K.-P., {Hurley}, J.~R., {Tout}, C.~A., \&
  {Eggleton}, P.~P. 1998, \mnras, 298, 525,
  \dodoi{10.1046/j.1365-8711.1998.01658.x}

\bibitem[{{Puls} {et~al.}(2006){Puls}, {Markova}, {Scuderi}, {Stanghellini},
  {Taranova}, {Burnley}, \& {Howarth}}]{2006A&A...454..625P}
{Puls}, J., {Markova}, N., {Scuderi}, S., {et~al.} 2006, \aap, 454, 625,
  \dodoi{10.1051/0004-6361:20065073}

\bibitem[{{Ram{\'\i}rez-Agudelo} {et~al.}(2013){Ram{\'\i}rez-Agudelo},
  {Sim{\'o}n-D{\'\i}az}, {Sana}, {de Koter}, {Sab{\'\i}n-Sanjul{\'\i}an}, {de
  Mink}, {Dufton}, {Gr{\"a}fener}, {Evans}, {Herrero}, {Langer}, {Lennon},
  {Ma{\'\i}z Apell{\'a}niz}, {Markova}, {Najarro}, {Puls}, {Taylor}, \&
  {Vink}}]{2013A&A...560A..29R}
{Ram{\'\i}rez-Agudelo}, O.~H., {Sim{\'o}n-D{\'\i}az}, S., {Sana}, H., {et~al.}
  2013, \aap, 560, A29, \dodoi{10.1051/0004-6361/201321986}

\bibitem[{{Ram{\'\i}rez-Agudelo} {et~al.}(2015){Ram{\'\i}rez-Agudelo}, {Sana},
  {de Mink}, {H{\'e}nault-Brunet}, {de Koter}, {Langer}, {Tramper},
  {Gr{\"a}fener}, {Evans}, {Vink}, {Dufton}, \& {Taylor}}]{2015A&A...580A..92R}
{Ram{\'\i}rez-Agudelo}, O.~H., {Sana}, H., {de Mink}, S.~E., {et~al.} 2015,
  \aap, 580, A92, \dodoi{10.1051/0004-6361/201425424}

\bibitem[{{Ram{\'\i}rez-Agudelo} {et~al.}(2017){Ram{\'\i}rez-Agudelo}, {Sana},
  {de Koter}, {Tramper}, {Grin}, {Schneider}, {Langer}, {Puls}, {Markova},
  {Bestenlehner}, {Castro}, {Crowther}, {Evans}, {Garc{\'\i}a}, {Gr{\"a}fener},
  {Herrero}, {van Kempen}, {Lennon}, {Ma{\'\i}z Apell{\'a}niz}, {Najarro},
  {Sab{\'\i}n-Sanjuli{\'a}n}, {Sim{\'o}n-D{\'\i}az}, {Taylor}, \&
  {Vink}}]{2017A&A...600A..81R}
{Ram{\'\i}rez-Agudelo}, O.~H., {Sana}, H., {de Koter}, A., {et~al.} 2017, \aap,
  600, A81, \dodoi{10.1051/0004-6361/201628914}

\bibitem[{{Riley} {et~al.}(2022){Riley}, {Agrawal}, {Barrett}, {Boyett},
  {Broekgaarden}, {Chattopadhyay}, {Gaebel}, {Gittins}, {Hirai}, {Howitt},
  {Justham}, {Khandelwal}, {Kummer}, {Lau}, {Mandel}, {de Mink}, {Neijssel},
  {Riley}, {van Son}, {Stevenson}, {Vigna-G{\'o}mez}, {Vinciguerra}, {Wagg},
  {Willcox}, \& {Team Compas}}]{2022ApJS..258...34R}
{Riley}, J., {Agrawal}, P., {Barrett}, J.~W., {et~al.} 2022, \apjs, 258, 34,
  \dodoi{10.3847/1538-4365/ac416c}

\bibitem[{{Ritter}(1988)}]{1988A&A...202...93R}
{Ritter}, H. 1988, \aap, 202, 93

\bibitem[{{Salpeter}(1955)}]{1955ApJ...121..161S}
{Salpeter}, E.~E. 1955, \apj, 121, 161, \dodoi{10.1086/145971}

\bibitem[{{Sana} {et~al.}(2012){Sana}, {de Mink}, {de Koter}, {Langer},
  {Evans}, {Gieles}, {Gosset}, {Izzard}, {Le Bouquin}, \&
  {Schneider}}]{2012Sci...337..444S}
{Sana}, H., {de Mink}, S.~E., {de Koter}, A., {et~al.} 2012, Science, 337, 444,
  \dodoi{10.1126/science.1223344}

\bibitem[{{Sana} {et~al.}(2013){Sana}, {de Koter}, {de Mink}, {Dunstall},
  {Evans}, {H{\'e}nault-Brunet}, {Ma{\'\i}z Apell{\'a}niz},
  {Ram{\'\i}rez-Agudelo}, {Taylor}, {Walborn}, {Clark}, {Crowther}, {Herrero},
  {Gieles}, {Langer}, {Lennon}, \& {Vink}}]{2013A&A...550A.107S}
{Sana}, H., {de Koter}, A., {de Mink}, S.~E., {et~al.} 2013, \aap, 550, A107,
  \dodoi{10.1051/0004-6361/201219621}

\bibitem[{{Sana} {et~al.}(2014){Sana}, {Le Bouquin}, {Lacour}, {Berger},
  {Duvert}, {Gauchet}, {Norris}, {Olofsson}, {Pickel}, {Zins}, {Absil}, {de
  Koter}, {Kratter}, {Schnurr}, \& {Zinnecker}}]{2014ApJS..215...15S}
{Sana}, H., {Le Bouquin}, J.~B., {Lacour}, S., {et~al.} 2014, \apjs, 215, 15,
  \dodoi{10.1088/0067-0049/215/1/15}

\bibitem[{{Sander} {et~al.}(2019){Sander}, {Hamann}, {Todt}, {Hainich},
  {Shenar}, {Ramachandran}, \& {Oskinova}}]{2019A&A...621A..92S}
{Sander}, A.~A.~C., {Hamann}, W.~R., {Todt}, H., {et~al.} 2019, \aap, 621, A92,
  \dodoi{10.1051/0004-6361/201833712}

\bibitem[{{Schneider} {et~al.}(2016){Schneider}, {Podsiadlowski}, {Langer},
  {Castro}, \& {Fossati}}]{2016MNRAS.457.2355S}
{Schneider}, F.~R.~N., {Podsiadlowski}, P., {Langer}, N., {Castro}, N., \&
  {Fossati}, L. 2016, \mnras, 457, 2355, \dodoi{10.1093/mnras/stw148}

\bibitem[{{Schnurr} {et~al.}(2008){Schnurr}, {Moffat}, {St-Louis}, {Morrell},
  \& {Guerrero}}]{2008MNRAS.389..806S}
{Schnurr}, O., {Moffat}, A.~F.~J., {St-Louis}, N., {Morrell}, N.~I., \&
  {Guerrero}, M.~A. 2008, \mnras, 389, 806,
  \dodoi{10.1111/j.1365-2966.2008.13584.x}

\bibitem[{{Shao} \& {Li}(2021)}]{2021ApJ...920...81S}
{Shao}, Y., \& {Li}, X.-D. 2021, \apj, 920, 81,
  \dodoi{10.3847/1538-4357/ac173e}

\bibitem[{{Shenar} {et~al.}(2019){Shenar}, {Sablowski}, {Hainich}, {Todt},
  {Moffat}, {Oskinova}, {Ramachandran}, {Sana}, {Sander}, {Schnurr},
  {St-Louis}, {Vanbeveren}, {G{\"o}tberg}, \& {Hamann}}]{2019A&A...627A.151S}
{Shenar}, T., {Sablowski}, D.~P., {Hainich}, R., {et~al.} 2019, \aap, 627,
  A151, \dodoi{10.1051/0004-6361/201935684}

\bibitem[{{Sibony} {et~al.}(2023){Sibony}, {Georgy}, {Ekstr{\"o}m}, \&
  {Meynet}}]{2023A&A...680A.101S}
{Sibony}, Y., {Georgy}, C., {Ekstr{\"o}m}, S., \& {Meynet}, G. 2023, \aap, 680,
  A101, \dodoi{10.1051/0004-6361/202346638}

\bibitem[{{Smith} \& {Maeder}(1991)}]{1991A&A...241...77S}
{Smith}, L.~F., \& {Maeder}, A. 1991, \aap, 241, 77

\bibitem[{{Stancliffe} \& {Eldridge}(2009)}]{2009MNRAS.396.1699S}
{Stancliffe}, R.~J., \& {Eldridge}, J.~J. 2009, \mnras, 396, 1699,
  \dodoi{10.1111/j.1365-2966.2009.14849.x}

\bibitem[{{Taam} \& {Sandquist}(2000)}]{2000ARA&A..38..113T}
{Taam}, R.~E., \& {Sandquist}, E.~L. 2000, \araa, 38, 113,
  \dodoi{10.1146/annurev.astro.38.1.113}

\bibitem[{{Tout} {et~al.}(1997){Tout}, {Aarseth}, {Pols}, \&
  {Eggleton}}]{1997MNRAS.291..732T}
{Tout}, C.~A., {Aarseth}, S.~J., {Pols}, O.~R., \& {Eggleton}, P.~P. 1997,
  \mnras, 291, 732, \dodoi{10.1093/mnras/291.4.732}

\bibitem[{{van den Heuvel}(1994)}]{1994inbi.conf..263V}
{van den Heuvel}, E.~P.~J. 1994, in Saas-Fee Advanced Course 22: Interacting
  Binaries, 263--474

\bibitem[{{van den Heuvel} {et~al.}(2017){van den Heuvel}, {Portegies Zwart},
  \& {de Mink}}]{2017MNRAS.471.4256V}
{van den Heuvel}, E.~P.~J., {Portegies Zwart}, S.~F., \& {de Mink}, S.~E. 2017,
  \mnras, 471, 4256, \dodoi{10.1093/mnras/stx1430}

\bibitem[{{van der Hucht}(2001)}]{2001NewAR..45..135V}
{van der Hucht}, K.~A. 2001, \nar, 45, 135,
  \dodoi{10.1016/S1387-6473(00)00112-3}

\bibitem[{{Vanbeveren} \& {Conti}(1980)}]{1980A&A....88..230V}
{Vanbeveren}, D., \& {Conti}, P.~S. 1980, \aap, 88, 230

\bibitem[{{Vanbeveren} {et~al.}(1998){Vanbeveren}, {De Loore}, \& {Van
  Rensbergen}}]{1998A&ARv...9...63V}
{Vanbeveren}, D., {De Loore}, C., \& {Van Rensbergen}, W. 1998, \aapr, 9, 63,
  \dodoi{10.1007/s001590050015}

\bibitem[{{Vink} {et~al.}(2001){Vink}, {de Koter}, \&
  {Lamers}}]{2001A&A...369..574V}
{Vink}, J.~S., {de Koter}, A., \& {Lamers}, H.~J.~G.~L.~M. 2001, \aap, 369,
  574, \dodoi{10.1051/0004-6361:20010127}

\bibitem[{{Wellstein} {et~al.}(2001){Wellstein}, {Langer}, \&
  {Braun}}]{2001A&A...369..939W}
{Wellstein}, S., {Langer}, N., \& {Braun}, H. 2001, \aap, 369, 939,
  \dodoi{10.1051/0004-6361:20010151}

\bibitem[{{Xu} \& {Li}(2010)}]{2010ApJ...716..114X}
{Xu}, X.-J., \& {Li}, X.-D. 2010, \apj, 716, 114,
  \dodoi{10.1088/0004-637X/716/1/114}

\bibitem[{{Yoon}(2017)}]{2017MNRAS.470.3970Y}
{Yoon}, S.-C. 2017, \mnras, 470, 3970, \dodoi{10.1093/mnras/stx1496}

\bibitem[{{Yoon} {et~al.}(2006){Yoon}, {Langer}, \&
  {Norman}}]{2006A&A...460..199Y}
{Yoon}, S.~C., {Langer}, N., \& {Norman}, C. 2006, \aap, 460, 199,
  \dodoi{10.1051/0004-6361:20065912}

\bibitem[{{Yu} {et~al.}(2019){Yu}, {Li}, {Zhu}, {Wang}, {Liu}, {Guo}, {Han},
  {Chen}, \& {L{\"u}}}]{2019ApJ...885...20Y}
{Yu}, J., {Li}, Z., {Zhu}, C., {et~al.} 2019, \apj, 885, 20,
  \dodoi{10.3847/1538-4357/ab44b5}

\bibitem[{{Yusof} {et~al.}(2013){Yusof}, {Hirschi}, {Meynet}, {Crowther},
  {Ekstr{\"o}m}, {Frischknecht}, {Georgy}, {Abu Kassim}, \&
  {Schnurr}}]{2013MNRAS.433.1114Y}
{Yusof}, N., {Hirschi}, R., {Meynet}, G., {et~al.} 2013, \mnras, 433, 1114,
  \dodoi{10.1093/mnras/stt794}

\bibitem[{{Zhang} \& {Jeffery}(2013)}]{2013MNRAS.430.2113Z}
{Zhang}, X., \& {Jeffery}, C.~S. 2013, \mnras, 430, 2113,
  \dodoi{10.1093/mnras/stt035}

\bibitem[{{Zhang} {et~al.}(2020){Zhang}, {Jeffery}, {Li}, \&
  {Bi}}]{2020ApJ...889...33Z}
{Zhang}, X., {Jeffery}, C.~S., {Li}, Y., \& {Bi}, S. 2020, \apj, 889, 33,
  \dodoi{10.3847/1538-4357/ab5e89}

\bibitem[{{Zhu} {et~al.}(2021){Zhu}, {Liu}, {Wang}, \&
  {L{\"u}}}]{2021A&A...654A..57Z}
{Zhu}, C., {Liu}, H., {Wang}, Z., \& {L{\"u}}, G. 2021, \aap, 654, A57,
  \dodoi{10.1051/0004-6361/202039692}

\bibitem[{{Zhu} {et~al.}(2023){Zhu}, {L{\"u}}, {Lu}, \&
  {He}}]{2023RAA....23b5021Z}
{Zhu}, C.-H., {L{\"u}}, G.-L., {Lu}, X.-Z., \& {He}, J. 2023, Research in
  Astronomy and Astrophysics, 23, 025021, \dodoi{10.1088/1674-4527/acafc7}

\end{thebibliography}
\bibliographystyle{./aasjournal}
\end{document}